\documentclass[11pt,prd,nofootinbib,preprint,superscriptaddress]{revtex4}

\usepackage[T1]{fontenc}
\usepackage{amsmath,amssymb, mathtools}
\usepackage{epsfig}
\usepackage{subfigure}
\usepackage{float}
\usepackage{graphicx}
\usepackage[usenames,dvipsnames]{color}
\usepackage{slashed}
\usepackage{multirow}
\usepackage{xcolor}
\usepackage[colorlinks,citecolor=blue]{hyperref}
\usepackage{pdfpages}
\usepackage{color}
\usepackage{comment}
\usepackage{soul}

\newcommand{\be}{\begin{equation}}
\newcommand{\ee}{\end{equation}}
\newcommand{\bea}{\begin{eqnarray}}
\newcommand{\eea}{\end{eqnarray}}

\newcommand{\bse}{\begin{subequations}}
\newcommand{\ese}{\end{subequations}}

\newcommand{\nn}{\nonumber}

\begin{document}

\title{EFT origin of Secluded Dark Matter}

\author{AseshKrishna Datta}
\email{asesh@hri.res.in}
\affiliation{Harish-Chandra Research Institute, A Constituent Institute of Homi Bhabha National Institute (HBNI), \\
	Chhatnag Road, Jhunsi, Prayagraj (Allahabad), Uttar Pradesh 211019, India}

\author{Sourov Roy}
\email{tpsr@iacs.res.in}
\affiliation{School of Physical Sciences, Indian Association 
	for the Cultivation of Science,\\  
	2A \& 2B Raja Subodh Chandra Mullick Road, Jadavpur, 
	Kolkata 700 032, India}

\author{Abhijit Kumar Saha}
\email{abhijit.saha@iopb.res.in}
\affiliation{Institute of Physics,
	  Sachivalaya Marg, Bhubaneswar 751005, Orissa, India\\ ~}

\author{Ananya Tapadar}
\email{intat@iacs.res.in}
\affiliation{School of Physical Sciences, Indian Association 
	for the Cultivation of Science,\\  
	2A \& 2B Raja Subodh Chandra Mullick Road, Jadavpur, 
	Kolkata 700 032, India}

\begin{abstract}
\vspace{2mm}
The present study aims to unveil a scenario with a non-minimal secluded dark sector (DS) in an effective field theory (EFT) framework. To explore this, we have examined a suitable extension of the type-X Two Higgs Doublet Model (2HDM) as a potential origin for the secluded DS. The DS comprises a dark matter (DM) candidate and a mediator particle `$a$' and possesses some non-minimal characteristics. It becomes
non-thermally populated through diverse dim-6 four-Fermi operators, effectively generated by integrating out the heavier Higgs particles. The analysis further focuses on the consequences of the collision processes $\textit{DM}+ a  \leftrightarrow a + a$ and $\textit{DM}+ \textit{DM}  \leftrightarrow a + a$ occurring within the DS. We have investigated the significance of employing an EFT approach in tracking the temperature evolution of the DS. Within the present framework, the observed relic abundance of the DM can be realized through both dark freeze-out and freeze-in mechanisms. Further, we have delineated the permissible ranges of the relevant parameters, viz., the DM mass ($m_\chi \gtrsim 20 \, \text{GeV}$), the portal coupling ($C_\tau \lesssim 10^{-14}\, \text{GeV}^{-2}$), and the DS coupling ($\lambda \lesssim 10^{-6} \,\text{GeV}^{-2}$) by taking into account the perturbativity of the involved couplings 
while reproducing the observed DM relic and complying with the bounds from a successful Big Bang Nucleosynthesis (BBN) and $\gamma$-ray searches.
\end{abstract}

\graphicspath{{figure/}}

\maketitle

\section{Introduction}
The existence of dark matter (DM) has now been firmly established through various astrophysical and cosmological observations~\cite{Sofue:2000jx, Clowe:2003tk, Clowe:2006eq, WMAP:2012nax, Planck:2018vyg}. However, the particle nature of the DM and its interactions with the Standard Model (SM) fields are still unknown. 
The absence of a viable candidate for the DM in the SM spectrum
has led several decades into intense exploration of potential physics scenarios beyond the SM (BSM) possessing a DM particle of varying nature. This has further prompted us to carry out detailed studies of plausible cosmological mechanisms that could generate the experimentally observed relic abundance of the DM.

In this context, one of the prominently debated topics aiming to elucidate the particle nature of the DM is the concept of Weakly Interacting Massive Particles (WIMPs)~\cite{Srednicki:1988ce, Gondolo:1990dk, Jungman:1995df, Feng:2008mu}. Under the WIMP paradigm, the DM  particles are expected to be in
thermal equilibrium with the SM particles in the early Universe and would decouple from the thermal bath later as they `freeze out'. The fact that such a weakly interacting DM particle, when having a mass in the electroweak scale, would possess a relic abundance matching the experimental measurements is commonly known as the ``WIMP miracle''~\cite{Feng:2010gw}. Despite being very appealing, such WIMPs have evaded detection so far. The absence of any convincing signature of the DM in direct and indirect searches severely restricts the viability of a WIMP DM~\cite{LUX:2015abn, SuperCDMS:2015eex, PandaX-II:2016vec, Roszkowski:2017nbc, Lin:2019uvt}. The null results of these searches might also hint at the possibility of new DM production mechanisms in thermal and non-thermal scenarios.
The most standard way of DM freeze-out in a conventional WIMP setup, following its decoupling from the thermal bath, typically involves an annihilation process like $DM \; DM \rightarrow a\,\, a$,  where `$a$' is any lighter particle present in the thermal bath. 

 Another mechanism for the freeze-out of thermal DM has been proposed in~\cite{ Kramer:2020sbb}, where the relic abundance is determined solely by the co-annihilation of the DM with such a lighter species `$a$', via the process $DM \, a \to a \, a$. Some other co-annihilation processes have been discussed in detail in references~\cite{Garny:2017rxs, Kim:2019udq, DAgnolo:2019zkf}.

However, it is possible that, in the early Universe, the DM could never reach a thermal equilibrium if it had very feeble interactions with particles in the thermal bath. Under such circumstances, the so-called freeze-in mechanism~\cite{Hall:2009bx, Bernal:2017kxu, Belanger:2023azf} could come into play which might still give rise to the observed DM relic abundance. Alternatively, it is also possible that a particle DM is an SM singlet and it resides inside a dark sector (DS) which remains thermally decoupled (secluded) from the visible sector (VS). However, a strong conversion process inside the DS allows the DM to reach an internal thermal equilibrium~\cite{Pappadopulo:2016pkp, Cheung:2010gj, Chu:2011be, hambye:2019, Evans:2019vxr, Du:2020avz, Krnjaic:2017tio, Berger:2018xyd, Ghosh:2022asg}. In that case, evolutions of the DS temperature ($T_D$) and the VS temperature ($T$) turn out to be significantly different~\cite{Pappadopulo:2016pkp, Cheung:2010gj, Evans:2019vxr, Du:2020avz, Krnjaic:2017tio, Ganguly:2022qxs}. The ratio $\frac{T_D}{T}$ crucially depends on how the DS particles were originally produced. In the case of a minimal secluded DS scenario, the DM depletes through annihilation to itself via DM number-changing self-interaction processes~\cite{Buen-Abad:2018mas,  Ghosh:2022asg}. In a scenario with a non-minimal secluded sector, the DM annihilates into lighter DS states via $\textit{DM} \,\, \textit{DM} \to a \, a$ scattering process where `$a$' (the mediator particle) is the lighter DS particle, and possesses feeble couplings to the VS particles
\footnote{The process, $DM \, a \to a \, a$, where `$a$' is the lighter DS species, has been discussed in reference~\cite{Berlin:2017ife}.}\cite{Pospelov:2007mp, Cheung:2010gj, Tapadar:2021kgw}.  

In this work, our primary aim is to develop an EFT description of a non-minimal secluded dark matter and subsequently examine its phenomenological ramifications\footnote{The EFT description of thermal WIMPs can be found in~\cite{Beltran:2008xg, Cheung:2012gi, Zheng:2010js, Belyaev:2018pqr, GAMBIT:2021rlp}.}. In particular, we investigate how the DS gets populated in the first place in a generic EFT-motivated framework where both $\textit{DM} \,\, \textit{DM} \; \leftrightarrow a \, a$, and $\textit{DM} \, a \leftrightarrow a \, a$ conversion processes could take place in a thermally decoupled or a secluded DS. Here, we have assumed that the DS is comprised of a single dark matter candidate $(\chi)$ and a lighter mediator particle ($\xi$; can be identified with `$a$' above), which are fermionic in nature. 

The population of the DS is originally induced and subsequently influenced by the VS through a series of dim-$6$ four-Fermi vertex operators. These operators are formulated within the framework of an EFT derived from a UV-complete model. The UV complete model we have considered for our purpose is an extended version of the type-X two Higgs Doublet Model (2HDM). The choice of a 2HDM scenario facilitates the incorporation of Yukawa-like portal interactions. On the other hand, our preference for only lepton-specific portal couplings makes us choose the type-X variant. The charged Higgs bosons of the type-X 2HDM would trigger lepton flavor universality violation (LFUV) at the tree level in the decays of the tau lepton into electron and muon and hence draws a lower bound on its mass~\cite{Logan:2009uf}.
Such an extended type-X 2HDM with lepton-specific portal interactions is further motivated by different astrophysical and cosmological observations~\cite{Nomura:2019wlo, Dey:2021alu}.

The particular choice of an extended 2HDM framework is solely to ensure generation of the required effective operators in the theory for DM masses ($m_\text{DM}$) ranging from $\mathcal{O}(10)\,\text{GeV}$ to $\mathcal{O}(10)\,\text{TeV}$. Such a scenario typically requires mass scales much heavier than $m_{\text{DM}}$, for example additional, heavy Higgs states, which are to be integrated out to generate the effective scale $\Lambda$. In passing we note that similar effective operators in the DS may also arise in more general or beyond–2HDM UV-frameworks, provided that sufficiently heavy Higgs fields are present in the theory.

Additionally, we have also incorporated a global $U(1)_{\mu - \tau}$ symmetry that gives rise to the portal couplings between $\xi$ and the $ \mu, \tau$ flavored leptons and also allows for both the conversion processes (as mentioned earlier) becoming active within the DS.
The $U(1)_{L_{\mu} - L_{\tau}}$ gauge theory, stands out as one of the anomaly-free gauge extensions, as discussed in Ref.~\cite{He:1990pn, He:1991qd}, and also in Ref.~\cite{ Ma:2001md} in the context of neutrino mass generation.
The global $U(1)_{\mu - \tau}$ symmetry is also well-motivated as it offers an explanation to the quasi-degenerate neutrino mass spectrum. 
At the leading order, this symmetry predicts the maximal atmospheric mixing (i.e., $\theta_{23} = 45^\circ$) and a vanishing reactor neutrino mixing angle, i.e., $\theta_{13}=0$, the solar angle ($\theta_{12}$) remaining unconstrained~\cite{Choubey:2004hn}. To reproduce neutrino oscillation data~\cite{ParticleDataGroup:2024cfk}, i.e., to obtain the observed nonzero values of $\theta_{12} $ and $ \theta_{13}$, the symmetry must be broken softly. It has been demonstrated in Refs.~\cite{Mohapatra:2004mf,Chamoun:2019pbh, Xing:2015fdg, Rodejohann:2017lre,  Xing:2022uax, Nath:2018hjx, Chauhan:2023faf, Liu:2017frs} that depending on different soft-breaking patterns, one can reproduce the observed neutrino mass-mixing.  
However, in the present work, we do not focus on the observed pattern of neutrino mass-mixing, as this would require inducting additional degrees of freedom beyond the minimal setup considered here. Detailed discussions on neutrino mass and mixing parameters in this kind of a scenario can be found elsewhere~\cite{Adhikary:2009kz, Rodejohann:2005ru}. These works reveal that, the additional right-handed neutrinos of such scenarios have to be rather heavy ($M_R \sim 10^{14-15}$ GeV), which is far above the effective scale $\Lambda$ considered in the present work. Hence, in a simple scenario that we propose in this work, we expect such a mechanism that generates the observed patterns in the neutrino mass-mixing would not affect our analysis which focuses on the DM sector.

Further, we assume the BSM Higgs bosons to be heavy enough such that those could be integrated out below the effective scale, $\Lambda$. This results in not only the relevant operators that induce the DS population in the first place but also those that would subsequently control the internal dynamics of the DS. To track the phases through which the DS passes as the Universe evolves, we further calculate the DS temperature, $T_D$, by considering the most relevant production modes of $\xi$, followed by the production of $\chi$ from $\xi$.
As the Universe cools down ($T_D < m_\chi$), $\chi$ annihilates to $\xi$ via the processes $\overline{\xi} \chi \to  \overline{\xi^c} \xi$ and $\overline{\chi}\chi \to \overline{\xi}\xi$ thus undergoing an eventual freeze-out, assuming the DM to be stable. 

In the scenario we propose in this work, the stability of the DM particle, $\chi$, is primarily aided by the usual consideration of a mass-hierarchy $m_\xi < m_\chi < 3 \, m_\xi$~\cite{Berlin:2017ife, Kramer:2020sbb}, which forbids its tree-level decay to $\xi$'s, i.e., $\chi \to \xi \xi \bar{\xi^c}$. It should, however, be noted that this arrangement may not suffice as the DM could then still undergo a loop-level decay, viz.,  $\chi \to \xi \bar{l} l$, which might threaten its stability unless the involved interactions are feeble enough. As long as one assumes a universal coupling parameter governing the interactions among the DS particles, its feebleness would immediately favor the freeze-in mechanism (over freeze-out) to set in (via the processes $\overline{\xi^c} \xi \to \overline{\xi} \chi$ and $\overline{\xi}\xi \to \overline{\chi}\chi$) that eventually would generate the observed relic abundance of a (stable) DM.

For a quantitative understanding, we study two representative situations. In the first case, both $\overline{\xi} \chi \leftrightarrow \overline{\xi^c} \xi$ and $\overline{\chi} \chi \leftrightarrow \overline{\xi} \xi$ processes are active within the DS while, in the second case, only the latter process is present. To estimate the DM relic abundance, we have constructed the set of coupled Boltzmann equations that govern the number density of the DS particles and the total energy density of the DS. We then solve them simultaneously in both the cases mentioned above. 

Furthermore, ensuring the smallness of the portal coupling that facilitates the initial non-thermal production/population of the DS is crucial to keeping the DS secluded. The structure of the proposed model allows for the decay of
$\xi$ into SM leptons at a later stage, via the same portal coupling, which might jeopardize the successful predictions of the Big Bang Nucleosynthesis (BBN). Both these considerations impose strong constraints on the VS\,-\,DS portal coupling. We duly consider these issues while looking for the viable region of the model parameter space that yields the correct DM relic abundance.
Further, the annihilation processes of the DM into a pair of mediator particles may give rise to a $\gamma$-ray spectrum which  
could be probed by the Fermi-LAT~\cite{Hooper:2010mq, Fermi-LAT:2012edv, Fermi-LAT:2015sau} telescope. 

The paper is organized as follows. In section~\ref{Model} we briefly describe the model under consideration and write down the relevant effective operators. Section~\ref{sec:Boltzmann_equation} and \ref{sec:study_of_BE} are devoted to a detailed analysis of the Boltzmann equations and discussions of our numerical results by complying with the theoretical requirements that the VS and the DS have to be out of mutual thermal equilibrium and that the involved couplings should remain perturbative. We also discuss constraints from the BBN and shed light on the prospects of indirect detection of the DM in $\gamma$-ray searches. We have further discussed that the constraint from LFUV on the mass of charged Higgs bosons  would set a maximum value of the effective scale of our theory. The summary and conclusions are presented in section~\ref{summary}. 
%
\section{The Model}
\label{Model}
In this study, we have investigated the type-X 2HDM model~\cite{Su:2009fz, PhysRevD.80.071701} (that constitutes the VS) by augmenting it with a DS comprised of an SM gauge singlet scalar $(S)$ and two Dirac Fermions: one acting as the mediator particle $(\xi)$ while the other $(\chi)$ proposed as the candidate for dark matter. We have introduced a global $U(1)_{\mu -\tau}$ symmetry, assigning non-trivial charges to $S$, $\xi$, and $\chi$, as detailed in Table~\,\ref{tab:charges}. The interactions that are relevant to this work are all contained in the Yukawa sector of the Lagrangian which are given as follows:
\begin{align}
\label{eq:lagrangian}
\mathcal{L} \supset y_{1i} \overline{L_i} \Phi_1 e_{R_i} + y_{2i} \overline{Q_i}\Phi_2 d_{R_i} + y_{3i} \overline{Q_i} \Phi_2^c u_{R_i}
+ y_4 \overline{L_\mu} \Phi_1^c \xi_R
+y_5 \overline{L_\tau} \Phi_1^c {\xi^c}_R+ y_6 S \overline{\xi^c} \xi + y_7 S \overline{\xi} \chi  + \textrm{h.c.} \, ,
\end{align}
where, $\Phi_1$ and $\Phi_2$ represent two distinct $SU(2)_L$ Higgs doublets. Notably, $\Phi_1$ exclusively couples to SM leptons, while $\Phi_2$ interacts exclusively with SM quarks. This coupling arrangement is specific to type-X 2HDM  and usually achieved by enforcing a $Z_2$ symmetry~\cite{Logan:2009uf}, wherein $\Phi_2$, along with the right-handed up-type quarks $u_{R_i}$ and right-handed down-type quarks $d_{R_i}$, exhibit odd parity under this symmetry. 
Once the electroweak symmetry breaks, the scenario gives rise to three Goldstone bosons that appear as the longitudinal modes of the massive $W^\pm$ and $Z$ bosons. Additionally, the spectrum contains five massive scalar degrees of freedom: two CP-even states $(h, \, H)$, one CP-odd state $(A)$, and a pair of charged Higgs states $(H^\pm)$. 
%
\begin{table}[h]
	\begin{center}
		\begin{tabular}{| c | c |}
			\hline
			Fields & $U(1)_{\mu - \tau}$  \\
			\hline
			$\xi$ & 1\\
			\hline
			$S$ & -2\\
			\hline
			$\chi$ & 3\\
			\hline
		\end{tabular}
		\caption{$U(1)_{\mu - \tau}$ charge assignments for the BSM gauge singlet fields.}
		\label{tab:charges}
	\end{center}
\end{table}
%
Here, both $\Phi_1$ and $\Phi_2$ acquire non-zero VEVs $v_1$ and $v_2$, respectively,  with, $v_{\text{SM}} = \sqrt{v_1^2 + v_2^2} = 246 \, \rm GeV$. The scalar doublets $\Phi_1$ and $\Phi_2$, in terms of the physical scalar states, can be written as,
\begin{align}
&\Phi_1=\begin{pmatrix}
-H^+\sin\beta\\
\sqrt{\frac{1}{2}}(v_1+H\cos\alpha-h\sin\alpha-i\,A\sin\beta)
\end{pmatrix}\,,\\
&\Phi_2=\begin{pmatrix}
H^+\cos\beta\\
\sqrt{\frac{1}{2}}(v_2+H\sin\alpha+h\cos\alpha+i\,A\cos\beta)
\end{pmatrix}\,,
\end{align}
where, $\tan\beta=\frac{v_2}{v_1}$ and $\alpha$ parameterize the mixing between the gauge and mass eigenstates of the CP-even Higgs bosons. In the alignment limit i.e. $\beta-\alpha=\frac{\pi}{2}$, `$h$' can be identified with the observed SM like Higgs boson having a mass $\simeq$ 125 GeV~\cite{Phalen:2006ga, Su:2009fz}. 

The Yukawa couplings $y_{1 i}$, $y_{2 i}$ and $y_{3 i}$ are given by,
\begin{align}
y_{1i} &=\frac{g \, m_{e_i}}{\sqrt{2} m_W\cos\beta},\,\, y_{2i}=\frac{g \, m_{u_i}}{\sqrt{2} m_W\sin\beta},\,\,y_{3i}=\frac{g \, m_{d_i}}{\sqrt{2} m_W\sin\beta},
\end{align}
where $e_i=\{e,\mu,\tau\}$, $u_i=\{u,c,t\}$ and $d_i=\{d,s,b\}$, include different generations of leptons, up-type and down type quarks, respectively and $m_{e_i}$,  $m_{u_i}$ and $m_{d_i}$ are the corresponding masses. The mass of the $W$ gauge boson is given by $m_W$, while `$g$' represents the $SU(2)_L$ gauge coupling. It may be noted that the most conservative upper limit on $\tan \beta$ comes from considering $y_{1\tau}< 4\pi$ and is given by,
\begin{align}
\tan\beta \lesssim 1255 \, .
\label{eq:tanbeta}
\end{align}

As for the interactions involving the DS particles, the existence of a global $U(1)_{\mu -\tau}$ symmetry dictates that only the fourth and fifth terms in Eq.~(\ref{eq:lagrangian}) are permissible as portal couplings linking the VS and the DS. Also for the same reason, the last two terms in Eq.~(\ref{eq:lagrangian}) are the only allowed ones that govern the interactions of DS particles among themselves. In subsequent sections, we will delve into the impact of these terms on the production mechanisms of the DS particles, the relic abundance of DM, and its detection prospects.

We impose a mass hierarchy of  $m_{S,H,A,H^\pm}$ $\gg$ $m_{\chi,\xi}$, which enables us to adopt an effective field theory framework with a cutoff scale $\Lambda$. The coupling of $\xi$ with the SM Higgs boson $(h)$ is suppressed by $\tan\beta$ and can be safely ignored for our purpose. To ensure a reasonably accurate approach to EFT at a high temperature, i.e., when the DS particles are relativistic to start with, we set $\Lambda = 100 \, m_\chi$~\cite{Bhattiprolu:2023akk}. Below this cutoff, on integrating out the heavier Higgs fields, several dim-6 four-Fermi operators emerge that contribute significantly to the production of the DS particles in the early Universe. The specific dim-6 operators are listed in Table~\ref{Table_2}.
\begin{table}[h]
	\begin{center}
		\begin{tabular}{ | c | } 
			\hline
			By integrating out $H^\pm$ \\ 
			\hline\\
			$\frac{C_{1\tau}}{m_{H^\pm}^2} \sin^2 \beta\, \overline{\nu_{\tau}}\tau_{R}\overline{\mu_{L}}\xi_R$\,,~~$\frac{C_{2\tau}}{m_{H^\pm}^2} \sin^2 \beta\, \overline{\nu_{\tau}}\tau_{R}\overline{\tau_{L}}\xi_{~R}^{c}$\,, ~~$\frac{C_3}{m_{H^\pm}^2} \sin^2 \beta\, \overline{\mu_L}\xi_R\overline{\xi^c_{~R}}\mu_L$ \\ 
			$\frac{C_4}{m_{H^\pm}^2} \sin^2 \beta\, \overline{\mu_L}\xi_R\overline{\xi_{R}}\mu_L$\,,~~$\frac{C_5}{m_{H^\pm}^2} \sin^2 \beta\, \overline{\tau_L}\xi^c_{~R}\overline{\xi^c_{~R}}\tau_L$\\
			\\
			\hline
			By integrating out $H$ and $A$\\
			\hline\\
			$\frac{C_{1\tau}}{m_{H}^2} \cos^2 \alpha\, \overline{\tau}\tau\overline{\nu_{\mu}}\xi_R$\,,~$\frac{C_{2\tau}}{m_{H}^2} \cos^2 \alpha\, \overline{\tau}\tau\overline{\nu_{\tau}}\xi_{~R}^c$\,,~$C_3\left(\frac{\cos^2\alpha}{m_{H}^2}- \frac{\sin^2\alpha}{m_{A}^2}\right) \overline{\nu_{\mu}}\xi_R\overline{\nu_{\tau}}\xi_{~R}^c$\,,\\
			$C_4\left(\frac{\cos^2\alpha}{m_{H}^2}+ \frac{\sin^2\alpha}{m_{A}^2}\right) \overline{\nu_{\mu}}\xi_R\overline{\nu_{\mu}}\xi_R$\,,~$C_5\left(\frac{\cos^2\alpha}{m_{H}^2}+ \frac{\sin^2\alpha}{m_{A}^2}\right) \overline{\nu_{\tau}}\xi^c_{~R}\overline{\nu_{\tau}}\xi^c_{~R}$\,,\\
			\\
			$\frac{C_{1\tau}}{m_A^2} \sin^2 \beta\, \overline{\tau}\gamma_5 \tau\overline{\nu_{\mu}}\xi_R$\,,~$\frac{C_{2\tau}}{m_A^2}\sin^2 \beta\, \overline{\tau}\gamma_5 \tau\overline{\nu_{\tau}}\xi_{~R}^c$\,,~$C_3\left(\frac{\cos^2\alpha}{m_{H}^2}- \frac{\sin^2\alpha}{m_{A}^2}\right)\, \overline{\xi_R}\nu_{\mu}\overline{\nu_{\tau}}\xi_{~R}^c$\,,\\
			$C_4\left(\frac{\cos^2\alpha}{m_{H}^2}- \frac{\sin^2\alpha}{m_{A}^2}\right) \overline{\nu_{\mu}}\xi_R\overline{\xi_R}\nu_{\mu}$\,,~$C_5\left(\frac{\cos^2\alpha}{m_{H}^2}- \frac{\sin^2\alpha}{m_{A}^2}\right) \overline{\nu_{\tau}}\xi^c_{~R}\overline{\xi_{~R}^c}\nu_{\tau}$\\
			\\
			\hline
            By integrating out $S$ \\
			\hline\\
            $\frac{\tilde{\lambda}_1}{m_S^2} \overline{\xi}\chi\overline{\xi}\xi^c \quad
            \frac{\tilde{\lambda}_2}{m_S^2} \overline{\chi}\chi\overline{\xi}\xi$ \\
            \\
            \hline
		\end{tabular}
         \caption{Effective four-Fermi operators obtained by integrating out the heavier Higgs bosons. The coefficients can be identified in terms of Yukawa interactions as $C_{1\tau} = y_{1\tau}y_4$, $C_{2\tau} = y_{1\tau}y_5$, $C_3 = y_4y_5$, $C_4 = y_4^2$ and $C_5 = y_5^2$, $\tilde{\lambda}_1 = y_6y_7$ and $\tilde{\lambda}_2 = y_7^2$.
         }	\end{center}
	\label{Table_2}
\end{table}
The Yukawa couplings $y_{1i}$ are fixed by the masses of the leptons for a given $\tan\beta$, whereas $y_4$, and $y_5$ are the free parameters of the theory.  Here, $y_{2i}$, $y_{3i}$ are also fixed by SM quark masses, `$i$' being the quark generation index.  Note that, $y_{2i}$, $y_{3i}$ do not contribute to the DS phenomenology since the quarks do not interact with the DS particles. It is further important to note that, the requirement of non-thermal production of the DS forces $y_4$ and $y_5$ to be very small as compared to $y_{1i},y_{2i}$ and which in turn makes $C_3$, $C_4$ and $C_5$ significantly suppressed. We also note that due to the known mass (Yukawa coupling) hierarchy of the SM leptons,
the primary contribution to the non-thermal production of the DS particles would predominantly come from terms involving the coefficient $C_{1\tau}$, $C_{2\tau}$. Since both $C_{1\tau}$ and $C_{2\tau}$ have similar roles to play in our analysis, for simplicity, we have assumed $C_{1\tau}= C_{2\tau}$. We further define 
\begin{equation}
\label{eq:ctau}
C_\tau=\frac{C_{1\tau}\sin^2\beta}{\Lambda^2}. 
\end{equation}
%
\begin{figure}[t]
    \centering
    \includegraphics[scale=1]{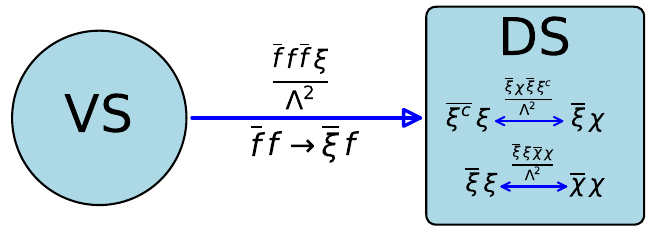}
    \caption{A cartoon illustrating the dynamics of DS phenomenology in the current setup. Both the initial population of the DS, as well as the conversion processes within the DS, are driven by dimension-6 four-Fermi operators, as indicated. Here, `$f$' represents the SM leptons.}
    \label{fig:Cartoon}
\end{figure}
%

Similarly, after integrating out `$S$' from the last two terms of Eq.~(\ref{eq:lagrangian}), we obtain two four-Fermi operators which are responsible for the conversion processes inside the DS\footnote{To develop the effective theory we have integrated out various mass scales at different stages, however for convenience in this work we have assumed the presence of a uniform cutoff scale considering the limit $\Lambda\simeq m_h, m_H^\pm, m_S$.} and hence for the time evolution of the number densities of the DS particles. These are 
\begin{subequations}\label{eq:convDS}
\begin{align}
\frac{\tilde{\lambda}_1}{m_S^2}\, \overline{\xi}\chi\overline{\xi}\xi^c\,~ \left(\text{for the process } \overline{\xi} \chi \leftrightarrow \overline{\xi^c} \xi \right),  \label{eq:zombie_type} \\
\frac{\tilde{\lambda}_2}{m_S^2} \,\overline{\chi}\chi\overline{\xi}\xi\,~ \left(\text{for the process } \overline{\chi} \chi \leftrightarrow \overline{\xi} \xi \right), \label{eq:wimp_type}
\end{align}
\end{subequations}
where $\tilde{\lambda}_1 = y_6y_7$ and $\tilde{\lambda}_2=y_7^2$, where $\tilde{\lambda}_1$ and $\tilde{\lambda}_2$ are the coupling parameters exclusive to the DS (see Table~\ref{Table_2}). A cartoon elucidating the complete dynamics of the DS in the present setup is presented in Fig.\,\ref{fig:Cartoon}.

Now, to understand the effect of the above-mentioned conversion processes on the DM production mechanism, we study the following two cases based on the relative magnitudes of $\tilde{\lambda}_1$ and $\tilde{\lambda}_2$:
\vskip 6pt
\noindent
{\large\textbf{\underline{Case-I:}}} 
This is a minimal scenario where, for simplicity, we assume $\tilde{\lambda}_1 = \tilde{\lambda}_2 = \tilde{\lambda}$
and we define a universal DS coupling parameter as $\lambda=\frac{\tilde{\lambda}}{\Lambda^2}$. For this, we note that the impact of the process $ \overline{\xi} \chi \to  \overline{\xi^c} \xi$ on DM freeze-out is significantly greater than that of the process $\overline{\chi}\chi \to \overline{\xi}\xi$. We discuss this further in section\,\ref{Analytic_understanding}. To ensure the stability of the DM, we first impose the kinematic condition $m_\xi < m_\chi< 3 m_\xi$ which prevents the tree level decay (3-body, $\chi \to \xi \xi \overline{\xi^c}$, driven by $\tilde{\lambda}_1$) of the DM. However, if $\tilde{\lambda}$ is critically large (and hence $\tilde{\lambda}_1$) the DM could still undergo the decay $\chi \to \xi \overline{l}l$ ($l \equiv \tau,\mu, \nu_\mu, \nu_\tau$) at one loop which also is driven by $\tilde{\lambda}_1$. 
Opting for a smaller $\tilde{\lambda}$, in turn, would clearly impede the freeze-out process. As we shall see later, in this case, freeze-in is the only viable mechanism that can explain the observed DM abundance. 

\vskip 6pt
\noindent
{\large \textbf{\underline{Case-II:}}}  One possible avenue to explore the viability of the DM freeze-out for obtaining the correct relic abundance is thus suppressing the problematic loop-mediated decay of the DM by making $\tilde{\lambda}_1$ sufficiently small ($\tilde{\lambda}_1 \ll \tilde{\lambda}_2$). An extreme situation of this case is when $\tilde{\lambda}_1 = 0$. In this case, the process $\overline{\chi}\chi \to \overline{\xi}\xi$  driven by Eq.~(\ref{eq:wimp_type}) would have to be the dominant one contributing to the DM freeze-out.  As in case-I here also remains the possibility of attaining the observed relic abundance via the freeze-in mechanism. 

In the subsequent sections, we discuss the dynamics of these cases in their analytical details with the help of Boltzmann equations while the corresponding numerical estimates are presented by adopting some benchmark points. The free parameters that control the DM phenomenology of the broad scenario are as follows:
\begin{align}
\{C_{\tau}, \lambda_1, \lambda_2, m_\xi, m_\chi\} \, ,
\end{align} 
 where, $\lambda_1 = \frac{\tilde{\lambda}_1}{\Lambda^2}$ and $\lambda_2 = \frac{\tilde{\lambda}_2}{\Lambda^2}$. An issue that is going to be important for our subsequent analyses is the perturbativity of the effective interaction strengths $C_\tau$ and $\lambda_{1,2}$ which sets upper bounds on their values. In this work, to meet this requirement, we demand
 \begin{align}
     \lambda_{1,2}, C_\tau < \frac{16 \pi^2}{(100 \, m_\chi)^2}\, ,
     \label{eq:perturbativity}
 \end{align}
which is consistent with (in fact, more stringent) what has been considered in the literature~\cite{Bhattiprolu:2023akk}.
%
\section{Studying the Boltzmann equations}
\label{sec:Boltzmann_equation}
In this section, we discuss in detail the Boltzmann equations that govern the time evolution of the number densities of the DS particles. 
We also present an analytical understanding of these equations with reference to the two DS processes indicated in Eqs. (\ref{eq:zombie_type}) and (\ref{eq:wimp_type}).
%
\subsection{Formulation of the Boltzmann equations}
\label{Boltzmann_equation}
As stated earlier, the DS is comprised of the fields $\xi$ and $\chi$ and decoupled from the VS bath. To start with, $\xi$ is produced non-thermally from the VS mainly via processes of the kind $\text{SM} + \text{SM} \to \xi + \text{SM}$. As 
discussed in section~\ref{Model}, the effective operators associated with the $\tau$-leptons play the dominant role in populating the DS. The contributing processes can be listed in two broad categories as follows: 
\begin{itemize}
\item those proceeding via operators obtained by integrating out heavy $H^\pm$:
\begin{align}
\label{eq:charged_higgs_process}
 & \overline{\tau_{_R}} \nu_\tau \to \overline{\mu_{_L}} \, \xi_R, \quad \overline{\tau_{_R}} \mu_{_L} \to \overline{\nu_\tau} \, \xi_R, \quad
    \mu_{_L} \nu_\tau \to \tau_{_R} \xi_R, \quad
    \overline{\mu_{_L}} \; \overline{\nu}_\tau \to \overline{\tau_{R}} \, \overline{\xi_R} \, , \quad (+ \, \text{h.c.}) \, , \nonumber
    \\ 
    & \overline{\tau_{_R}} \nu_\tau \to \overline{\tau_{_L}} \, \xi^c_R, \quad
    \overline{\tau_{_R}} \tau_{_{L}} \to \overline{\nu}_\tau \, \xi^c_R, \quad
    \tau_{_{L}} \nu_\tau \to \tau_{_{R}} \, \xi^c_R, \quad
    \overline{\tau_{_{L}}} \, \overline{\nu}_\tau \to \overline{\tau_{_R}} \, \overline{\xi^c_R} \,, \quad 
    (+ \, \text{h.c.}) \, ,
 \end{align}
%
\item those proceeding via operators obtained by integrating out heavy $H, \, A$: 
 \begin{align}
\label{eq:neutral_higgs_process}
  &\overline{\tau} \tau \to \overline{\nu}_\mu \, \xi_R, \quad
  \overline{\tau} \nu_\mu \to \overline{\tau} \, \xi_R, \quad
  \tau \nu_\mu \to \tau \, \xi_R, \quad
  \overline{\tau} \, \overline{\nu}_\mu \to \overline{\tau} \, \overline{\xi_R} \, , \quad (+ \, \text{h.c.}) \, , \nonumber
  \\
  & \overline{\tau} \tau \to \overline{\nu}_\tau \, \xi^c_R, \quad
    \overline{\tau} \nu_\tau \to \overline{\tau} \, \xi^c_R, \quad
    \tau \nu_\tau \to \tau \, \xi^c_R, \quad
    \overline{\tau} \, \overline{\nu}_\tau \to \overline{\tau} \, \overline{\xi^c_R}, \quad (+ \, \text{h.c.}) \,.
 \end{align}
%
\end{itemize}

For an optimally strong mutual interaction between $\xi$ and $\chi$, and both having sufficient number densities, these two states could reach a local (within DS) kinetic equilibrium with unique temperature
$T_D\neq T$~\cite{Heeba:2018wtf,Hryczuk:2021qtz}, where $T$ corresponds to the 
temperature of the VS. As discussed in section~\ref{Model}, the proximity in 
masses of these two states makes it imperative to track their number 
densities ($n_\chi$ and $n_\xi$) simultaneously. In addition, since the dark sector particles 
have a temperature different from that of the VS ones, one should also consider 
the evolution of the dark sector energy density ($\rho_{D}=\rho_\chi+\rho_\xi$) as 
a function of $T$. The relevant set of Boltzmann equations
with the collision term (CE) appearing on the right-hand side (RHS) are as follows: 
%
\begin{eqnarray}       
 \frac{dn_{\chi_{\text{tot}}}}{d t}+3 \mathcal{H} n_{\chi_{\text{tot}}} &=& \frac{1}{2}\langle\sigma v\rangle_{\overline{\xi^c} \xi\to \overline{\xi} \chi}^{T_D}\left[n_{\xi_{\text{tot}}}^2-n_{\chi_{\text{tot}}} n_{\xi_\text{tot}} \frac{n_{\xi_{\text{tot}}}^{\text{eq}}(T_D)}{n_{\chi_{\text{tot}}}^{\text{eq}}(T_D)}\right] \nonumber \\
&+& \frac{1}{2}\langle\sigma v\rangle_{\overline{\xi} \xi \to \overline{\chi}\chi}^{T_D}\left[n_{\xi_{\text{tot}}}^2-n_{\chi_{\text{tot}}}^2 \frac{n_{\xi_{\text{tot}}}^{\rm eq}(T_D)^2}{n_{\chi_{\text{tot}}}^{\rm eq}(T_D)^2}\right]\,,
\label{eq:Beq_chi}
\\  \nonumber \\ \nonumber
 \frac{dn_{\xi_{\text{tot}}}}{dt} +\frac{dn_{\chi_{\text{tot}}}}{d t} + 3 \mathcal{H} (n_{\xi_{\text{tot}}} +n_{\chi_{\text{tot}}} ) &=& 
\frac{1}{2} \left[\langle\Gamma_{\xi \to \rm SM}\rangle^Tn_{\xi_{\text{tot}}}^{\rm eq}(T) - \langle\Gamma_{\xi \to \rm SM}\rangle^{T_D} n_{\xi_{\text{tot}}} \right] \nonumber \\
&+& 2 \gamma_{_{\text{SM},\,\text{SM} \to \text{SM},\, \xi}}(T)\,,
\label{eq:Beq_xi}
\\ \nonumber \\
\dfrac{d\rho_{D_{\text{tot}}}}{dt} +3 \mathcal{H} (\rho_{D_{\text{tot}}}+p_{D_{\text{tot}}}) &=& \frac{1}{2}\Gamma_{\xi \to\text{SM}} \, m_\xi (n_{\xi_{\text{tot}}}^{\rm eq}(T) - n_{\xi_{\text{tot}}}) + 2 \Upsilon_{\text{SM},\,\text{SM} \to \text{SM}\xi}(T)\, , \nonumber \\
\hfill 
\label{eq:Beq_energy}
\end{eqnarray}
%
where $n_{i_{\text{tot}}} = n_{\overline{i}} + n_{i} $ is the total number density of the $i^{\rm th}$ particle with,
$n_i^{\rm eq}(T) = \frac{g_i}{2 \pi^2} m_i^2 \, T \,K_2(\frac{m_\chi}{T})$, $\rho_{D_\text{tot}} =  \rho_{\overline{\chi}} + \rho_{\chi}  + \rho_{\overline{\xi}} + \rho_{\xi}$ and $p_{D_\text{tot}} =  p_{\overline{\chi}} +p_{\chi}   + p_{\overline{\xi}}+ p_{\xi}$
are the total energy density and pressure of the DS, respectively. In a 
relativistic regime of the DS particles, the equation of state is characterized by $p = \frac{1}{3} \rho$, while during matter-domination, it simplifies to
$p = 0$. For simplicity, we have assumed the Maxwell-Boltzmann statistics for 
both  the $\chi$ and  $\xi$ and express $\rho_i$ and $p_i$ (where $i=\{\chi,\xi\}$) in  the following form:
\begin{align}
\rho_i &= n_i \left(3 T_D + m_i \dfrac{K_1\left(\dfrac{m_i}{T_D}\right)}{K_2\left(\dfrac{m_i}{T_D}\right)}\right) \, , \\
p_i &= n_i T_D \,,
\end{align}
where, $m_i$ is the mass of the $i^{th}$ species and $K_1$ and $K_2$ are the 
modified Bessel functions of second kind. While deriving the Boltzmann 
equations, we have considered a matter-antimatter symmetry within the DS. This 
symmetry explains the presence of the $\frac{1}{2}$ and $2$ factors on the right-hand side 
(RHS) of Eqs.~(\ref{eq:Beq_chi}), (\ref{eq:Beq_xi}) and (\ref{eq:Beq_energy}). 
Once the evolution of $\rho_D$ as a function of $T$ is known, it is possible to 
extract the evolution of $T_D$ with $T$.

In Eq.~(\ref{eq:Beq_chi}) and Eq.~(\ref{eq:Beq_xi}),
$\langle\sigma v\rangle_{\overline{\xi^c} \xi \to \overline{\xi}\chi}^{T_D}$ and 
$\langle\sigma v\rangle_{\overline{\xi}\xi\to \overline{\chi}\chi}^{T_D}$ 
represent the thermally averaged cross-sections for the processes
$\overline{\xi^c}\xi\to \overline{\xi}\chi$ and $\overline{\xi}\xi\to \overline{\chi}\chi$, respectively, at temperature $T_D$. In Eq.~(\ref{eq:Beq_xi}) $\langle\Gamma_{\xi \to \rm SM}\rangle^T = \Gamma_{\xi \to \rm SM} \frac{K_1\left(\frac{m_\xi}{T}\right)}{K_2\left(\frac{m_\xi}{T}\right)}$ refers to the 
thermal average of three-body decay width of $\xi$ into the VS particles at temperature 
$T$, where $\Gamma_{\xi \to \rm SM}$ is the total decay width of $\xi$. The terms 
$\gamma_{_{\text{SM},\,\text{SM} \to \text{SM},\, \xi}}(T)$ in 
Eq.~(\ref{eq:Beq_xi}) and $\Upsilon_{\text{SM},\,\text{SM} \to \text{SM},\xi}$ in 
Eq.~(\ref{eq:Beq_energy}) correspond to the collision terms contributing to the 
production component (arising from the VS) feeding the number density and the energy density of the DS particles, respectively. The expressions for these terms 
are as follows:
\begin{align}
\gamma_{ab \to c\xi}(T) &=  \dfrac{8 \pi^2 g_a g_b T}{(2 \pi)^6} \int_{s_{\rm min}}^\infty ds \, \lambda(s,m_c,m_\xi)^2 s^{3/2} \sigma_{c\xi \to ab}(s) K_1({\sqrt{s}/T)} \\
\Upsilon_{ab \to c\xi}(T) &=  \dfrac{4 \pi^2 g_c g_d T}{(2 \pi)^6} \int_{s_{\rm min}}^\infty ds \, \lambda(s,m_c,m_\xi)^2 s^2 \left(1-R(s,m_c,m_\xi)\right) \sigma_{c\xi \to ab}(s) K_2({\sqrt{s}/T)}
\end{align}
Here, $\lambda(s,m_a,m_b) = \frac{1}{2 s}\left[(s-(m_a+m_b)^2)( s-(m_a-m_b)^2)\right]^{1/2}$ and $s_{\rm min} = \mathrm{min}[(m_a+m_b)^2, (m_c+m_d)^2]$ 
and $R(s,m_c,m_d) = \frac{m_d^2-m_c^2}{s}$. Similarly, the two terms with positive 
coefficients in the RHS of Eq.~(\ref{eq:Beq_energy}) serve as the source terms for 
the DS energy density with origins in inverse decay and scattering of SM 
particles, respectively, at early Universe. Due to the smallness of $T_D$, we have 
safely ignored the backward reaction processes
$(\xi + \text{SM} \to \text{SM} + \text{SM})$ in the Boltzmann equation. The terms 
with a negative coefficient in the respective RHS of Eq.~(\ref{eq:Beq_xi}) and Eq.~(\ref{eq:Beq_energy}) induce the fall of $n_{\xi_{\rm tot}}$ and
$\rho_{D_{\rm tot}}$ due to late time decay of $\xi$. The inclusion of the decay 
term for $\xi$ in Eq.~(\ref{eq:Beq_xi}) also ensures that $\xi$ is not sufficiently 
abundant in the present-day Universe thus establishing the present setup as a single 
component DM framework. 
%
\subsection{An analytic understanding of the Boltzmann equations}
\label{Analytic_understanding}
In this subsection, we aim to analytically comprehend the Boltzmann equations presented in Eqs.~(\ref{eq:Beq_chi}), (\ref{eq:Beq_xi}), and (\ref{eq:Beq_energy}). We study the time-evolution of the DS temperature ($T_D$) and the roles played by the two key conversion processes (as indicated in Eqs.~(\ref{eq:zombie_type}) and (\ref{eq:wimp_type})) that determine the relic abundance of the DM.  

To begin with, let us concentrate on Eq.~(\ref{eq:Beq_energy}) which governs the 
evolution of the energy density of the DS. We assume that $\chi$ and $\xi$ are in kinetic
equilibrium after the initial non-thermal production of $\xi$ from the VS.
If the coupling within the DS is not enough, $\chi$ and $\xi$ will not be in chemical equilibrium. Such a situation can open up the possibility of producing DM via a freeze-in mechanism from $\xi$'s produced initially from the VS.
However, if the interactions within the DS are strong enough, there would be a chemical equilibrium between $\chi$'s and $\xi$'s. Driven by the existence of various processes within the DS, once a chemical equilibrium gets established, the principle of detailed balance would dictate $\mu_\chi  =  \mu_\xi = \mu_D$, where $\mu_\chi$ and $\mu_\xi$ are the effective chemical potentials of $\chi$'s and $\xi$'s, respectively. The expression for $\mu_D$ is
\begin{align}
\label{eq:effchempot}
\mu_D &= T_D \log \left(\frac{n_i}{n_i^{\text{eq}}}\right).
\end{align}
The effect of non-zero chemical potential on the DM freeze-out has been discussed in Ref.~\cite{ Bandyopadhyay:2011qm, Farina:2016llk, Berlin:2017ife, Kramer:2020sbb}. Consequently, in such a scenario, the relativistic DS particles will have an effective equilibrium number density $(n^{\rm eq, eff})$ which for the DM $\chi$ is given by 
\begin{align}
\label{Eq:effEqDM}
n_\chi^{\rm eq, eff}(T) &= \dfrac{n_\xi (T)}{n_\xi^{\rm eq} (T_D)}n_\chi^{\rm eq}(T_D) \nonumber \\ 
&=  n_\chi^{\rm eq}(T_D) \, e^{\frac{\mu_D}{T_D}}\,.
\end{align}
To obtain an analytical estimate of $T_D$, we assume that  $\mu_D$  is significantly smaller than the energies of the respective particles and hence can be neglected. Therefore, when the DS particles are relativistic in nature, we can write $\rho_D = \frac{\pi^2}{30} g_{D_\rho} T_D^4$, where $g_{D_\rho}=g_\chi+g_\xi$ with $g_\chi$ and $g_\xi$ being the internal degrees of freedom for $\chi$ and $\xi$, respectively\footnote{For finite values of $\mu_D$, $\rho_D$ can be expressed as $\rho_D = \frac{\pi^2} {30} g_{D_\rho} T_D^4 e^{\frac{\mu_D}{T_D}}$.}. It turns out that the impact of the decay term on the RHS of Eq.~(\ref{eq:Beq_energy}) on the production of $\xi$ at early times is subdominant compared to that of the scattering term and hence can be safely ignored. We thus write $\sigma_{\xi \text{SM} \to  \text{SM},\,\text{SM}}\simeq \frac{C_\tau^2}{16 \pi } s$ in the limit $\sqrt{s}\gg m_{\chi,\xi}$ where, $C_\tau$ is the effective coupling of $\xi$ with the VS particles. With these, the collision term (involving $2\to 2$ process) appearing in the RHS of Eq.~(\ref{eq:Beq_energy}) can be estimated as,  
\begin{align}
\Upsilon_{\text{SM},\,\text{SM} \to \text{SM}\xi}^{\text{appx}}(T) &= \dfrac{4 \pi^2 g_a g_b T}{(2 \pi)^6} \int_{s_{\rm min}}^\infty ds \, \lambda(s,m_a,m_b)^2 s^2 \frac{C_\tau^2}{16 \pi } s \, K_2({\sqrt{s}/T)} \,, \nonumber \\
&= C_\tau^2 \dfrac{24}{\pi^5} T^9
\label{eq:collision_energy}
\end{align}
Next, we derive an approximate analytical expression for $T_D$ (a similar approach has been adopted in ref.\,\cite{Evans:2019vxr}) given by
\begin{align}
     \zeta(T) &= \left[ \int_{T_i = \Lambda}^{T} dT^\prime \dfrac{30 \Upsilon_{\text{SM},\,\text{SM} \to \text{SM}\xi}^{\text{appx}}(T^\prime)}{\pi^2 g_{*D} \mathcal{H}(T^\prime)T^{\prime^5}} \right]^{1/4},
     \label{eq:temp_ratio_analytic}
\end{align}
where $\zeta = \frac{T_D}{T}$ and $T_i$ corresponds to the cutoff scale $\Lambda$ of the EFT. If we consider the Universe to be radiation-dominated at the time of production of the DS particles, then by using Eq.~(\ref{eq:collision_energy}) and Eq.~(\ref{eq:temp_ratio_analytic}) we find
\begin{align}
 \label{eq:Td_n0_anlaytic}
 T_D &= \sqrt{C_\tau} T  \left(\dfrac{80 \, Mp \Lambda^3}{14.94 \, \pi^7 g_{D*} \sqrt{g_\rho}} \right)^{1/4} \propto \sqrt{C_\tau} T \Lambda^{3/4}\left(M_P \right)^{1/4} \,.
\end{align}
Thus, in the early Universe, when the DS particles are relativistic, $T_D$ turns out to be proportional to $T$. Such a correlation has also been reported in ref.~\cite{hambye:2019}. 
On the other hand, if the DS is induced by some renormalizable couplings,  
$T_D$ does not redshift as radiation in the early Universe 
\cite{Evans:2019vxr,Tapadar:2021kgw,Ganguly:2022qxs} even when all the DS particles are 
relativistic.

Note that $\mu_D$ may not always be negligible 
(as assumed till now). If that is the case, i.e., for $\mu_D \neq 0$, one also has to consider the evolution of $\frac{\mu_D}{T_D}$ with temperature $(T)$. Under the circumstances, to obtain an accurate description of the DS dynamics, one requires solving the coupled set of 
Boltzmann equations. Nevertheless, the relationship between $T_D$ and 
$T$, as mentioned earlier (see Eq.~(\ref{eq:Td_n0_anlaytic}), remains valid and this can be gleaned from the numerical results presented in the next section (section~\ref{sec:study_of_BE}).

We now move on to the set of Boltzmann equations in Eqs.~(\ref{eq:Beq_chi}) and (\ref{eq:Beq_xi}). Given that $m_\chi \approx m_\xi$, these two equations are to be solved simultaneously.
For simplicity, let us assume $\mu_D = 0$ and thus consider the analytical expressions for $T_D$ in Eq.~(\ref{eq:Td_n0_anlaytic}). Here, both the terms on 
the RHS of Eq.~(\ref{eq:Beq_chi}) would 
contribute to the production of DM ($\chi$). Once the DS 
reaches thermal equilibrium with a temperature $T_D$, the DS 
particles attain the effective equilibrium number 
density~\cite{Kramer:2020sbb, Berlin:2017ife}. As $T_D$ 
drops below $m_\chi$, DM $(\chi)$ starts annihilating to $\xi$ via $\overline{\xi}\chi \to \overline{\xi^c} \xi$ and $\overline{\chi} \chi \to \overline{\xi} \xi$. 
Consequently, the number density of $\xi$ ($n_\xi$) starts increasing thus prompting an eventual ``freeze-in'' for these particles. 
 On the other hand, the number density of DM $(\chi)$ decreases, while approaching a 
freeze-out. 
It is important to note that due to non-renormalizable couplings, both $\langle\sigma_{\chi \overline{\chi} \to \xi \overline{\xi}} \rangle, \langle\sigma_{\overline{\xi}\chi\to  \overline{\xi^c}\xi} \rangle \propto C_\tau^2 m_\chi^2$.
Notably, the contributions from both the annihilation modes to the freeze-out process, at the time of DM freeze-out (characterized by the freeze-out temperature $T_{D_{\text{fo}}}$) can be understood by comparing the reaction rates of these processes, i.e.,
\begin{align}
\frac{\Gamma_{ \overline{\xi}\chi\to \overline{\xi^c}\xi}}{\Gamma_{ \overline{\chi} \chi \to  \overline{\xi}\xi}} &= \dfrac{n_\xi (T_{D_{\text{fo}}}) }{n_\chi (T_{D_{\text{fo}}}) } \dfrac{ \langle \sigma_{\overline{\xi}\chi\to  \overline{\xi^c}\xi}\rangle} {\langle \sigma_{ \overline{\chi} \chi \to \overline{\xi}\xi } \rangle}\,.
\label{eq:rate-ratio}
\end{align}
Given that the effective operators that control these processes have similar strengths, we may consider $\frac{\langle\sigma_{\chi \overline{\chi} \to \xi \overline{\xi}} \rangle}{\langle\sigma_{\chi\overline{\xi^c}\to  \chi\overline{\xi}}\rangle} \simeq 1$. Then, by using the condition for the so-called sudden freeze-out, we can arrive at $n_\chi (T_{D_{\text{fo}}}) = n_\chi^{\text{eq}} (T_{D_{\text{fo}}})$. On the other hand, 
a freeze-in production of $\xi$ (during the DM freeze-out) implies
$n_\xi(T_{D_{\text{fo}}}) > n_{\xi_{\text{eq}}}(T_{D_{\text{fo}}})$.
Eq.~(\ref{eq:rate-ratio}) then leads to the following relation:
\begin{align}
\frac{\Gamma_{ \overline{\xi}\chi\to \overline{\xi^c}\xi}}{\Gamma_{ \overline{\chi} \chi \to  \overline{\xi}\xi}} &>
  \dfrac{n_\xi^{\text{eq}} (T_{D_{\text{fo}}}) }{n_\chi^{\text{eq}} (T_{D_{\text{fo}}}) }  \,, \nn \\
     &>  \left(\dfrac{m_\xi}{m_\chi} \right)^{3/2} \exp \left({\frac{m_\chi - m_\xi}{m_\chi}\, x_{D_\text{fo}}}\right)\,,
\end{align}
where, $x_{D_{\text{fo}}} = \frac{m_\chi}{T_{D_{\text{fo}}}} \sim 20$. This implies that, in our scenario, the reaction rate of
$ \overline{\xi} \chi\to \overline{\xi^c}\xi$  would be much larger than that of $\overline{\chi} \chi \to  \overline{\xi} \xi$ at the time of DM freeze-out.

As mentioned in section~\ref{Model}, in Case-I both the DS processes are active, out of which the process $\overline{\xi} \chi\to \overline{\xi^c}\xi$ plays the dominant role in determining the DM relic abundance via the freeze-out mechanism. However, later we check that the freeze-out process for Case-I is ruled out by the stability of the DM. On the other hand, for feeble DS couplings, the correct DM relic abundance comes from the freeze-in production of the DM $\chi$ from $\xi$. In this situation  both the processes $\overline{\xi^c} \xi \to \overline{\xi} \chi$ and $\overline{\xi}\xi \to \overline{\chi}\chi$ become relevant.  In Case-II since only one DS process (as indicated in Eq.~(\ref{eq:wimp_type})) is active, the DM relic abundance in both freeze-out (large DS coupling) and freeze-in (small DS coupling) scenarios is governed by this process. 
%
\section{Results}
\label{sec:study_of_BE}
In this section, we discuss various constraints on the 
parameter space of the broad scenario under consideration that 
come from the theory and relevant experiments. The former set includes the one originating in the requirement of the VS and the DS to be out of a thermal equilibrium and the others concerning the perturbativity of the (effective) coupling parameters $\lambda_i$'s and $C_\tau$'s. The latter set stems from the demand for achieving a successful BBN, from compliance with the photon flux observed by an experiment like the Fermi-LAT in its bid to find indirect signatures of the DM and also from the requirement of satisfying the LFUV constraint.
We carry out a thorough study of the scenario by numerically solving the Boltzmann equations presented in section~\ref{Boltzmann_equation} and estimate the DM relic abundance for both the cases introduced in section~\ref{Model}.
%
\subsection{Regions (out) of thermal equilibrium and of a perturbative \texorpdfstring{$C_\tau$}{C tau}}
\label{Thermal_equilibrium_region}
We identify the region of parameter space where the VS and the DS are out of 
thermal equilibrium by first delineating the region where these sectors are indeed 
in thermal equilibrium. Towards this, we follow the conventional approach that 
involves the computation of the collective reaction rates responsible for the 
production of the particles belonging to the VS from the DS ones, juxtaposing 
those with the Hubble parameter ($\mathcal{H}$). The quantity that determines 
whether a thermal equilibrium is established or not between the two sectors is 
$\frac{\Gamma_{\xi \, \text{SM} \to \text{SM} \, \text{SM}}}{\mathcal{H}}$, where 
$\frac{\Gamma_{\xi \, \text{SM} \to \text{SM} \, \text{SM}}}{\mathcal{H}} > (<) 1$ 
implies the equilibrium is (not) realized.
\begin{figure}[t]
        \begin{center}
        \includegraphics[scale=0.6]{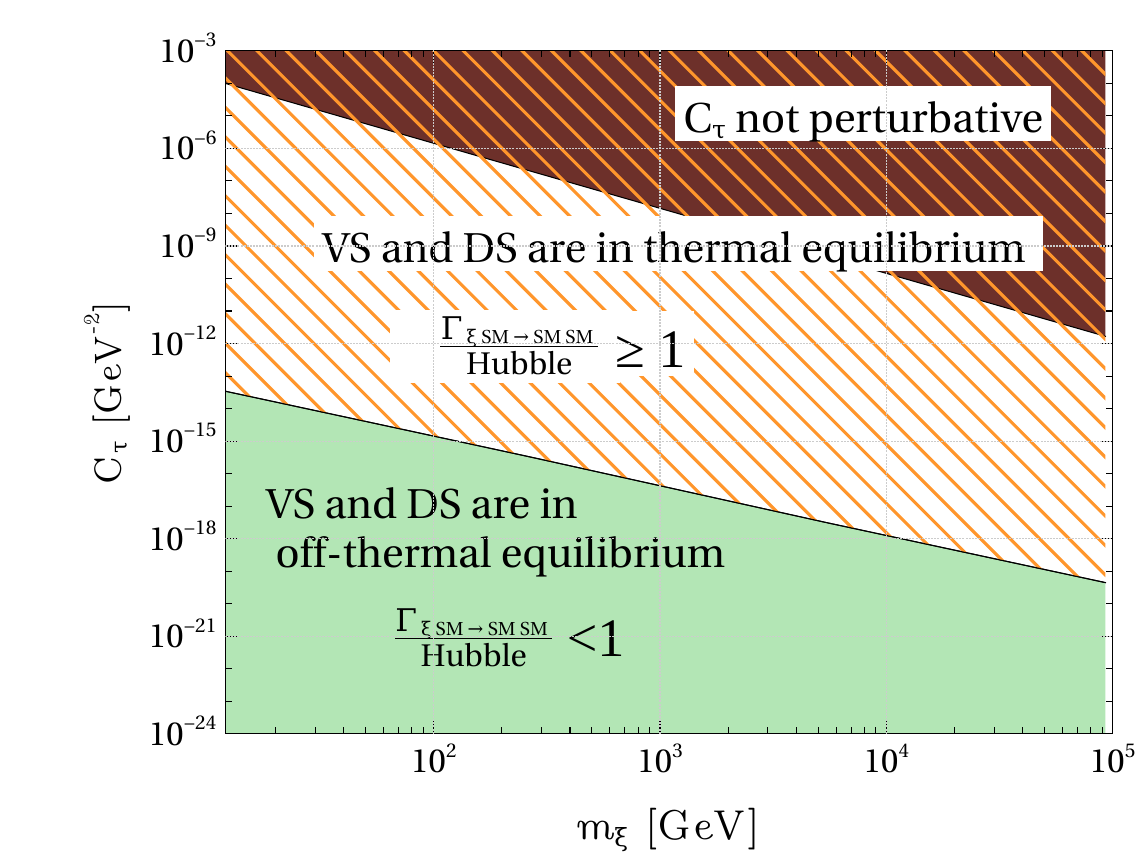}
		\caption{Situation with a possible thermal equilibrium between the VS and the DS presented in the $m_\xi-C_\tau$ plane with $T=\Lambda
\, (= \, 100 \, m_\chi)$. Also shown is the region where $C_\tau$ ceases to be perturbative. See text for details.}
		\label{fig:Thermal_equilibrium}
        \end{center}
\end{figure}

In Fig.~\ref{fig:Thermal_equilibrium}, we contrast the regions in the
$m_\xi-C_\tau$ plane over which the VS and the DS could eventually move into a 
thermal equilibrium as opposed to where those remain out of equilibrium, the latter 
being the target region of the present study. The choice of the indicated plane is 
prompted by the fact that $m_\xi$ and $C_\tau$ are the two parameters that 
primarily control the ratio
$\frac{\Gamma_{\xi \text{SM} \to \text{SM} \, \text{SM}}}{\mathcal{H}}$, 
the very quantity that decides, in the first place, the feasibility of thermal 
equilibrium. In addition, $m_\xi$ restricts the DM mass, $m_\chi$, such that the 
latter not only aids its stability but could also account for its observed value of 
relic abundance. As for $C_\tau$, it plays a role in accounting for
the observed relic abundance and ensuring a successful BBN. Furthermore, the figure is drawn for $T=\Lambda \, (=100 \, m_\chi)$. The choice of the cutoff scale, $\Lambda$, restricts the value of $C_\tau$ from above (see Eq.~\ref{eq:perturbativity}), beyond which the theory violets perturbativity of $C_\tau$. The corresponding region is shown by the
brown, wedge-shaped area at the top part of the figure.  

The hashed strip in orange illustrates the region where the VS and the DS coexist in thermal 
equilibrium. Conversely, the coveted region over which the two 
sectors would be out of thermal equilibrium is shown in light green.  
Notably, as  $m_\xi$ increases, the reaction rates get enhanced; we thus looks 
for smaller $C_\tau$'s for larger values of $m_\chi$ so that the two sectors 
still remain out of thermal equilibrium.

At this point, note that the interactions between the VS and the DS 
are controlled by the dim-6 effective operators, as discussed in 
section~\ref{Model}. The processes that describe the production of the VS states 
from the DS ones are just the reverse ones of those mentioned in 
Eqs.~(\ref{eq:charged_higgs_process}) and (\ref{eq:neutral_higgs_process}). The expression for the relevant collision term at very high temperatures can be 
obtained in the same way as has been for $\gamma_{ab \to c\xi}$ in 
section~\ref{Analytic_understanding} and is found to vary as
$\gamma_{ab \to c\xi} \propto C_\tau^2 T^8$. Consequently, the reaction rates for the processes  
$\xi \, \text{SM} \to \text{SM}\, \text{SM}$ are given by
$\Gamma_{ \xi  \text{SM}\to \text{SM} \, \text{SM}} \propto C_\tau^2 T^5$. 
Thus, the quantity $\frac{\Gamma_{\xi \text{SM} \to \text{SM} \, \text{SM}}}{\mathcal{H}}$ 
drops as $T$ decreases. This signifies that a DS of a secluded kind at a higher 
temperature would remain so at a lower temperature. 

Thus, a departure from a thermal equilibrium between the VS and the DS may 
be allowed at a higher value of $T \, (> 100 \, m_\chi)$ and hence such a possibility remains germane 
for the present work. However, this necessitates a smaller $C_\tau$ to throw the 
VS and the DS systems out of mutual equilibrium at that high temperature. As we will see later
(see Fig.~\ref{fig:scan}), such a smaller value of $C_\tau$ tends to 
attract constraints from BBN. Thus, these two requirements, in conjunction, restrict the value of $C_\tau$ from both above and below in our scenario with a secluded DS. It is also found that the requirement of the VS and the DS to be off-equilibrium sets an upper bound on $C_\tau$ more stringently than what the requirement of perturbativity of the same does. 
%
\subsection{Constraints from the BBN}
\label{BBN}
In the standard picture of BBN, around temperature $\lesssim\mathcal{O}(1)$ MeV, the energy of the cosmic microwave background (CMB) photons becomes lower than the binding energies of the light elements, e.g., deuterium (D), $^3$He, $^4$He, $^7$Li and hence these elements/isotopes get synthesized~\cite{Tytler:2000qf}. There are strict observational constraints on the primordial abundances of these light elements (see section 2 of reference~\cite{Kawasaki:2020qxm} and references therein). If a long-lived particle is produced in the early Universe and it decays subsequently into SM particles at a late time, the BBN predictions might get altered. Thus, given that the relevant observables have now been precisely measured, such an effect can be used to constrain the model parameters.

In our framework, the particle $\xi$ is unstable and can be long-lived depending on the smallness of the coupling coefficient $C_{\tau}$. Once the DS freezes out or freezes in, $\xi$ finally decays to the SM leptons. If such decays occur during the BBN, the newly produced charged leptons could emit energetic photons (electromagnetic shower) which, in turn, could interact with the cosmic microwave background radiation (CMBR) and trigger the photo-dissociation process~\cite{Jedamzik:2006xz, Kawasaki:2017bqm, Kawasaki:2020qxm, Alves:2023jlo} thus altering the abundance of light elements. On the other hand, high-energy neutrinos that might appear from such decays could scatter off the background leptons and produce charged leptons and charged pions. The former could, once again, induce electromagnetic showers that might destroy light elements that are already synthesized while the latter impacts the $n-p$ ratio through nucleon and pion interactions~\cite{Kanzaki:2007pd}. Hence to evade any significant impact on the outcome of BBN,
it is apparently safe to make $\xi$ decay early enough with  $\tau_\xi<1 \, \text{second}$, which translates to $T>3$~MeV. The constraint on the relevant model parameters ($m_\xi$ and $C_\tau$) from BBN will be presented later in section~\ref{sec:final_overview}.
%
\subsection{The DM relic abundance}
\label{sec:numerical_analysis}
The study of the DM relic abundance is entirely based on the analysis of the
Boltzmann equations. Thus, in this section, we first thoroughly explore the implications of these coupled set of equations (presented in 
Eqs.~(\ref{eq:Beq_chi}), (\ref{eq:Beq_xi}) and (\ref{eq:Beq_energy})) for the DS processes by solving them numerically.
Towards this, we track the evolution of effective chemical potential $\mu_D$ (discussed in section~\ref{sec:Boltzmann_equation}) and the number densities ($n_{\chi,\xi}$) of $\chi$ and $\xi$ as functions of $x = \frac{m_\chi}{T}$. We do this by choosing some benchmark points for the two cases (Case-I and Case-II) that are introduced in section~\ref{Model}.
Throughout, as mentioned in section~\ref{Model}, we set the
cutoff scale of the effective theory to be $\Lambda = 100 \, m_\chi$.
%
\subsubsection{Case-I (\texorpdfstring{$\lambda_1 = \lambda_2 = \lambda$}{lambda1 = lambda2 = lambda})} 
\label{sec:case1} 
In this case, we consider the Yukawa couplings related to the DS to be of equal strength. Consequently, we have $\frac{\tilde{\lambda}_1}{\Lambda^2}=\frac{\tilde{\lambda}_2}{\Lambda^2} = \lambda$ whose value is to be chosen suitably. 
To have a quantitative understanding of the DS dynamics, we consider a benchmark point (BP-1) consisting of three remaining free parameters as given by
\begin{align}
\text{BP-1:} \,  \left\{C_{\tau},m_\xi, m_\chi  \right\} &\equiv \left\{1.5 \times 10^{-16} \, \text{GeV}^{-2},\, 10^2\, \text{GeV}, 1.1 \times 10^2 \, \text{GeV}  \right\}. 
\end{align}
The choice of such a benchmark point is guided primarily by the stability of the DM and compatibility with the BBN constraint which, as we have discussed earlier, are in some tension.
\begin{figure}[t]
    \begin{center}
		\includegraphics[scale = 0.5]{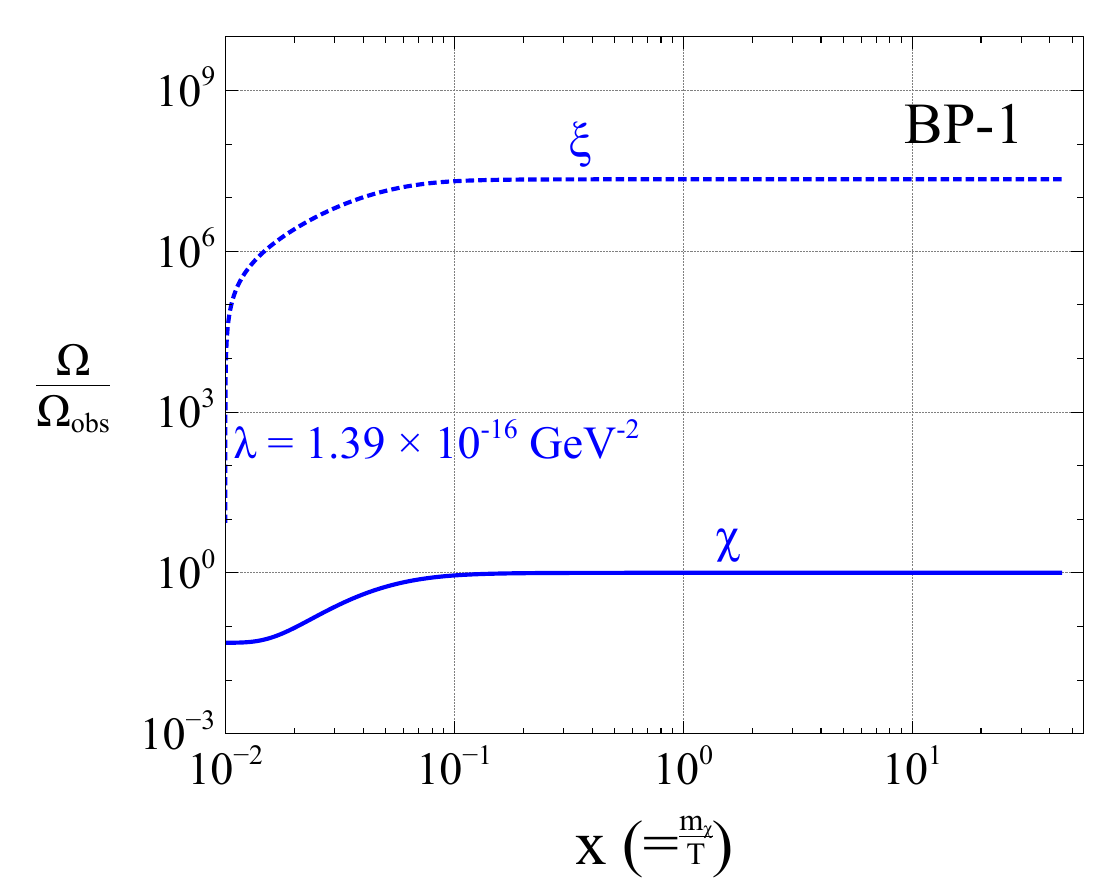} 
		\caption{Variations of $\Omega_i/\Omega_{\text{obs}}$ with respect to $x = m_\chi/T$ for the scenario BP-1 under Case-I. The value of $\lambda$ for which the observed DM relic abundance is reproduced is indicated. Here, the solid (dashed) lines represent evolutions of $\chi$ ($\xi$).
        See text for details.}
		\label{fig:line_plot_1_temp}
    \end{center}
\end{figure}

In Fig.~\ref{fig:line_plot_1_temp}, we discuss the temperature evolutions of the number densities for $\chi$ and $\xi$ ($n_\chi$ and $n_\xi$, respectively) for benchmark point BP-1 in terms of more conventional quantities $\Omega_i/\Omega_{\text{obs}}$,  where, the abundance $\Omega_{i= \chi, \xi} = \frac{m_\chi s_0 Y_i}{\rho_{\text{crit}}} =2.755 \times 10^8 \left(\frac
{m_i}{\rm \, 1 GeV}\right) Y_i(T)$, with $Y_i (T) = \frac{n_i(T)}{s(T)}$, $s(T)$ being the entropy density of the Universe while $s_0$ indicates its present value. $\Omega_{\text{obs}}$, on the other hand, is the observed DM relic abundance. Such evolution for $\chi$ and $\xi$ are illustrated with solid and dashed lines, respectively, and the  and the value of $\lambda$ for which the DM relic abundance can be achieved within
its experimentally reported range is also indicated. We stick to these line-type conventions throughout the rest of the paper to represent the evolution of number densities of $\chi$ and $\xi$ (in terms of $\Omega_i/\Omega_{\text{obs}}$ ) with $x$.

We consider a value of $\lambda$ ($= 1.39 \times 10^{-16}\, \text{GeV}^{-2}$) in Fig~\ref{fig:line_plot_1_temp} where the corresponding variations for $\Omega_i/\Omega_{obs}|_{i=\chi,\xi}$ 
are indicated in blue colors.
For such a small value of $\lambda$, the DS is unable to reach an internal thermal equilibrium and both the processes $\overline{\xi^c}\xi \to \overline{\xi} \chi$ and $\overline{\xi} \xi \to \overline{\chi}\chi$ contribute in attaining the final (observed) relic abundance of the DM,  via the freeze-in mechanism. Although, in our analysis, we respect the kinematic condition $m_\xi < m_\chi <3m_\xi$ that prohibits the primary (tree-level) 3-body decay of the DM ($\chi \to \xi \xi \overline{\xi^c}$), the loop-mediated decay of the DM, $\chi \to \xi \overline{l}l$, could pose a serious threat to the stability of the DM. However, for the above choice of parameters, the DM is found to be stable with its lifetime ($\tau_\chi \simeq 3.3 \times 10^{29} \, \text{sec}$) far exceeding the age of the Universe ($\simeq 10^{17}$ sec) and also exceeding its lower bound coming from $\gamma$-ray searches($\mathcal{O} \left(10^{17} \right)$ sec, see references~\cite{Ibarra:2007wg, Slatyer:2016qyl, Dutta:2022wuc}). 

We have further checked that for a much larger value of $\lambda$ (in this case $\lambda = 1.89 \times 10^{-8}\, \text{GeV}^{-2}$), a chemical equilibrium between $\chi$ and $\xi$ can set in and as the Universe evolves, i.e., the temperature drops, $\chi$ starts annihilating to $\xi$ via the DS processes $\overline{\xi} \chi \to \overline{\xi^c}\xi$ and $\overline{\chi} \chi \to \overline{\xi} \xi$. Hence the number density of $\chi$ starts decreasing and consequently, the DM seems to freeze out at the observed relic abundance ballpark. However, at a later time, the DM will decay into $\xi$ via $\chi \to \xi \overline{l}l$ at loop level (as the tree-level decay is forbidden by the aforesaid kinematic condition). The lifetime of the DM in this case turns out to be $\tau_\chi \simeq 1.94 \times 10^{13} \, \text{sec}$ which makes the DM decay too rapidly. Thus, the only viable mechanism that could give rise to the observed DM relic abundance in Case-I is the one of the freeze-in type. The freeze-out mechanism is found to have a serious tension with the stability of the DM.

It should be noted that the small value of $\lambda$ that is required to provide a stable DM while possibly ensuring a correct DM relic abundance, falls short of the present-day sensitivity of not only the direct but also the indirect (DM) detection experiments. The issue is discussed in further detail in section~\ref{indirectSearch}. 
%
\subsubsection{Case-II (\texorpdfstring{$\lambda_1 \ll \lambda_2$)}{lambda1 << lambda2}}
\label{sec:case-II}
In this section, we first investigate if the freeze-out mechanism could result in the observed DM relic abundance while keeping the DM stable. We note that the DM decay width ($\Gamma_{\chi \to \xi \overline{l}l}$) varies as $\lambda_1^2$. Hence a possible way to find a stable DM is by considering a suitably smaller value of $\lambda_1$ (as mentioned in section~\ref{Model}) such that $\lambda_1 \ll \lambda_2$.  
 Consequently, the DS dynamics is now dominated by $\lambda_2$ which is the Wilson coefficient for the four-Fermi operator $\overline{\chi}\chi \overline{\xi}\xi$. We further look for a viable freeze-in mechanism consistent with all relevant constraints. 

 With such a hierarchy between $\lambda_1$ and $\lambda_2$,
 to carry out a quantitative study of the DS dynamics, we start with the benchmark point BP-1 which is characterized by relatively small $m_{\xi,\chi} \sim 100 \, \text{GeV}$. 
 As we shall soon find, any lower value of $m_\xi$ (and hence $m_\chi$) quickly attracts constraints from the BBN (see Fig.~\ref{fig:scan}) in the secluded DS scenario.
 This prompts us to explore a much heavier DM which inevitably requires a smaller $\lambda_2$ due to the perturbativity constraint that we discuss by referring to Fig.~\ref{fig:scan}. The corresponding benchmark point that we choose is given by
 \begin{align}
 \text{BP-2: } \lbrace C_\tau, m_\xi, m_\chi \rbrace \equiv \left\{ 10^{-21} \, \text{GeV}^{-2},\, 10 \, \text{TeV}, 11 \, \text{TeV}  \right\}\,, 
 \end{align}
 which still maintains the same ratio between $m_\chi$, and $m_\xi$ as in the case of BP-1. The point BP-2 is characterized by pretty large values of $m_\chi$ and $m_\xi$ and hence will be consistent only with rather small values of $\lambda_2$. As discussed earlier, we solve the coupled set of Boltzmann equations presented in Eqs.~(\ref{eq:Beq_chi}), (\ref{eq:Beq_xi}) and (\ref{eq:Beq_energy}).  For the sake of simplicity, we set $\lambda_1 = 0$ and work with finite values of $\lambda_2$. We then look into the evolutions of $\frac{\mu_D}{T_D}$, $T_D$ and the number densities of $\chi$ and $\xi$, i.e.,  $n_\chi$ and $n_\xi$.

\begin{figure}[t]
    \begin{center}
        \includegraphics[height=6cm,width=7.5cm]{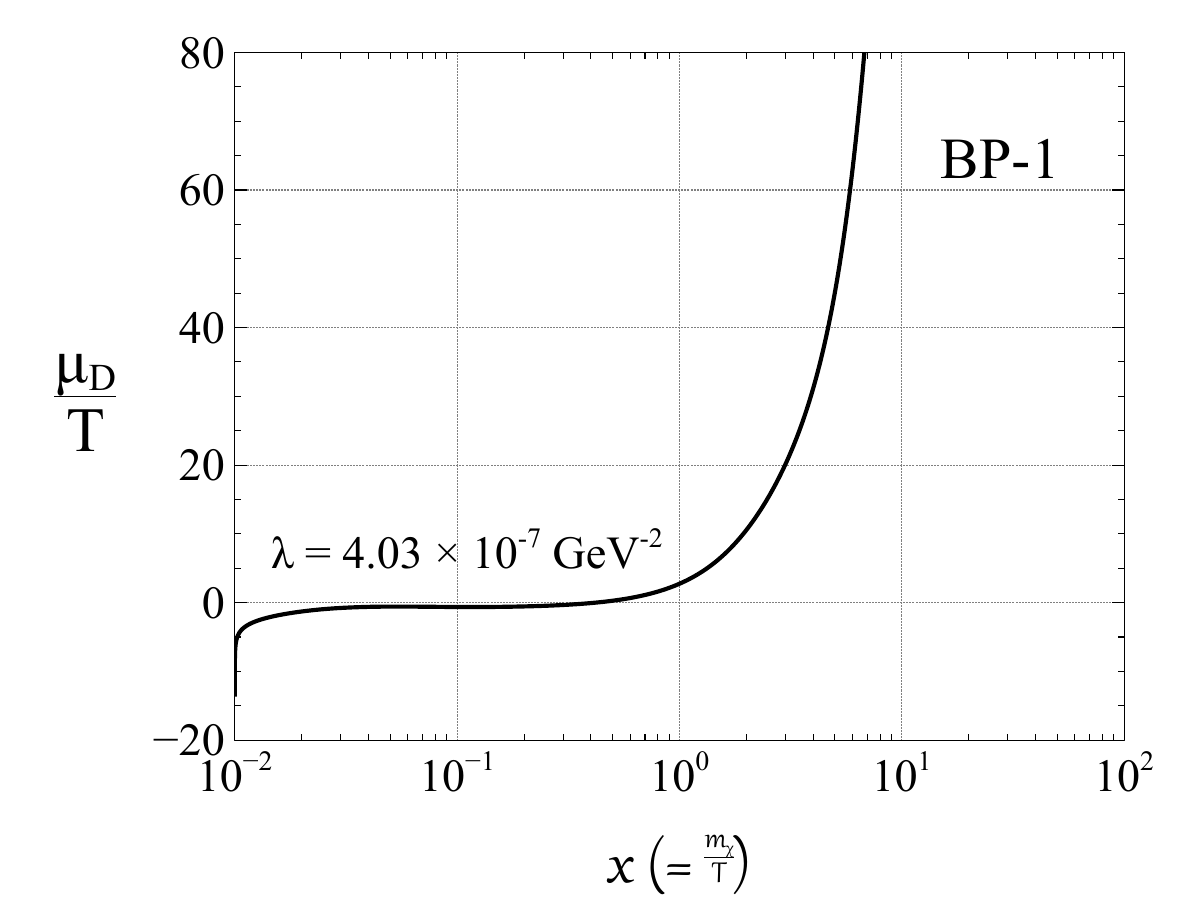}
        \hspace{1cm}
		\includegraphics[height=6cm,width=7.5cm]{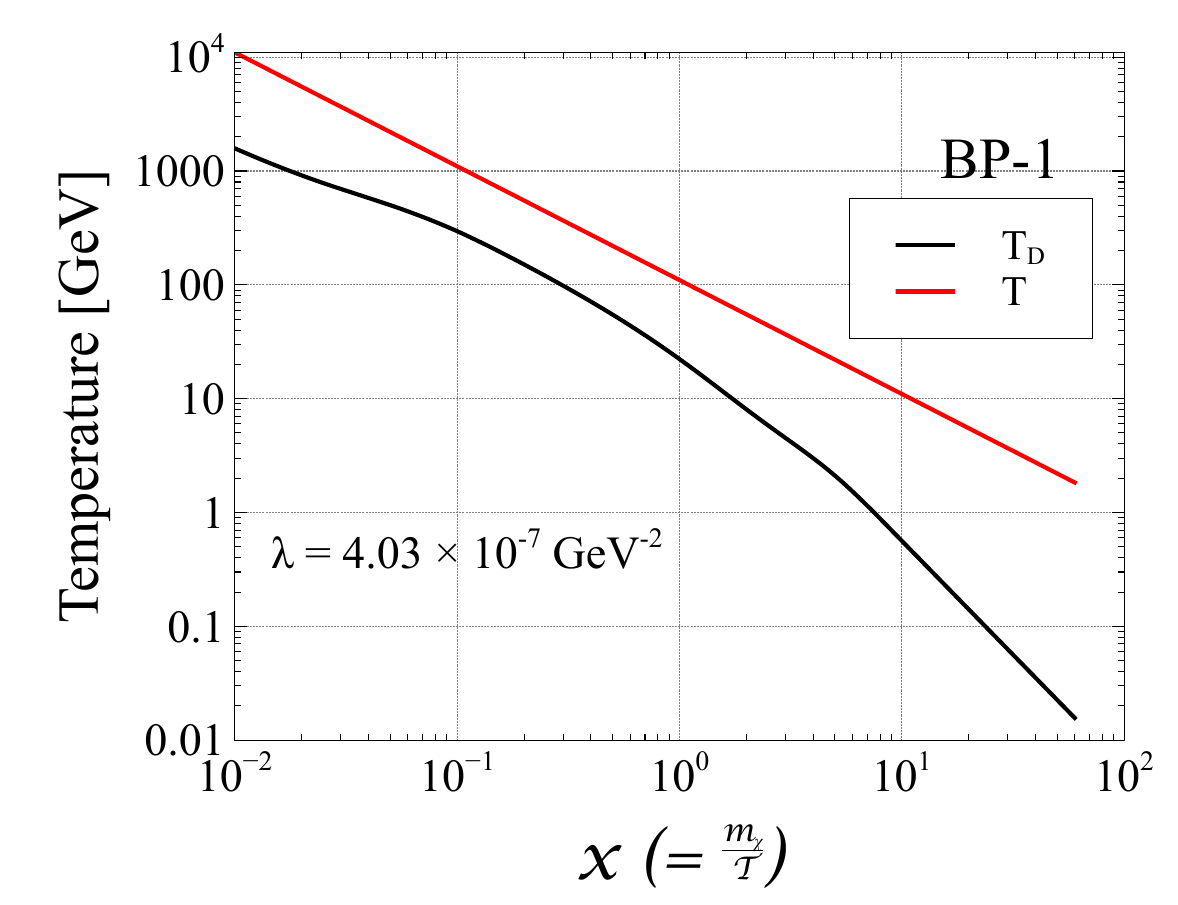} 
		\caption{Variations of $\frac{\mu_D}{T_D}$ (left) and $T_D$ and $T$ (right; in black and red, respectively) as functions of $x (=m_\chi/T)$ for the scenario BP-1 under Case-II. The appropriate value of $\lambda_2$ that results in the correct relic abundance of DM is also indicated. See text for details.}
		\label{fig:temp_case_II}
    \end{center}
\end{figure}

As pointed out in section~\ref{sec:Boltzmann_equation}, we work with the most general situation having $\mu_D \neq 0$. 
In the left plot of Fig.~~\ref{fig:temp_case_II}, we present the variation of $\frac{\mu_D}{T_D}$ with `$x$' for the benchmark scenario BP-1. The corresponding value of $\lambda_2$ that is used for the purpose is also indicated. The value is so chosen that it reproduces the DM relic abundance within the range reported by experiments. 
It can be seen, that as the temperature ($T$) of the Universe drops (when the DM freezes out; see the right plot), the ratio $\frac{\mu_D}{T_D}$ starts increasing. This is what is expected from Eq.~(\ref{eq:effchempot}).

In the right plot of Fig.~\,\ref{fig:temp_case_II}, we now choose to show the evolutions of $T_D$ (in black) and $T$ (in red) as functions of `$x$'. 
It is seen that, while $T$ drops with increasing `$x$' (as defined) for a given $m_\chi$, $T_D$ initially increases sharply because of a transfer of energy from the VS bath to the DS as dictated by the set of Boltzmann equations presented in section~\ref{sec:Boltzmann_equation}. As the Universe cools down (i.e., $T$ decreases), $T_D$ drops as well.  For $T_D < m_\xi$ (i.e., $x > 1$), when both the DS particles, $\chi$ and $\xi$, turn non-relativistic, $T_D$ starts red-shifting like matter.

One can have a similar set of plots for the scenario BP-2
that would have aided in understanding a possible freeze-out of
DM in the said scenario. However, we do not present those as we have checked the value of $\lambda_2$ ($3.27 \times 10^{-9}\, \text{GeV}^{-2}$) that is required for yielding the right DM relic (within 1$\sigma$) would be ruled out by the perturbativity constraint thus turning the freeze-out mode no more viable for achieving the correct DM relic.  

We now turn to the discussion of evolutions of the number densities $n_\chi$ and $n_\xi$ in terms of the ratio, $\Omega_i/\Omega_{\text{obs}}|_{i=\chi, \xi}$ (as introduced in the previous subsection) and check if those could result in the observed DM relic following the freeze-out (red lines) and/or freeze-in (blue lines) path(s). For freeze-out to take place, one needs an optimally large $\lambda_2$ such that there exists a chemical equilibrium between the DS states $\chi$ and $\xi$. Subsequently, when the DS temperature $T_D$ becomes comparable to $m_\chi$, $\chi$ starts annihilating to $\xi$ via the process $\overline{\chi} \chi \to \overline{\xi}\xi$.  
Finally, when the reaction rate of this particular process drops below $\mathcal{H}$, the DM freezes out. Note that the other annihilation mode $\overline{\xi} \chi \to \overline{\xi^c}\xi$ is now closed as we have set $\lambda_1= 0$.

\begin{figure}[t]
    \centering
    \includegraphics[height=6cm,width=7.5cm]{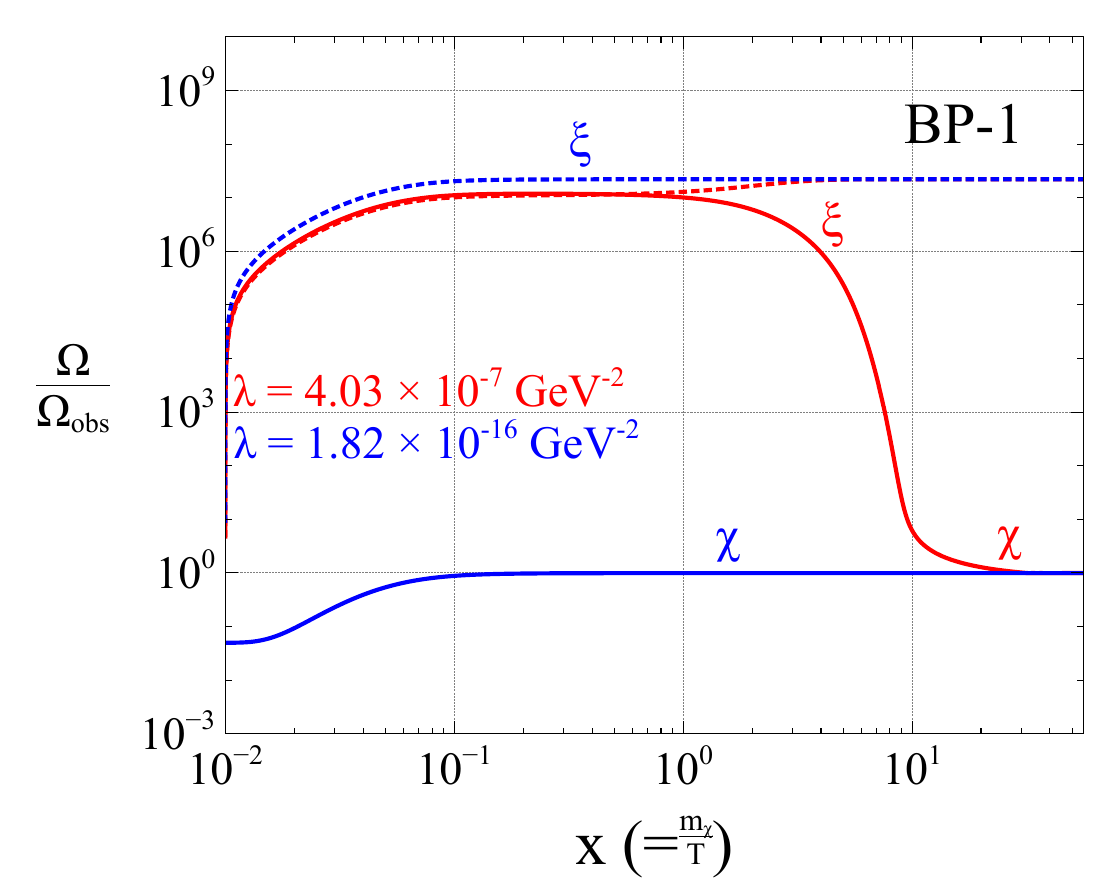}
    \includegraphics[height=6cm,width=7.5cm]{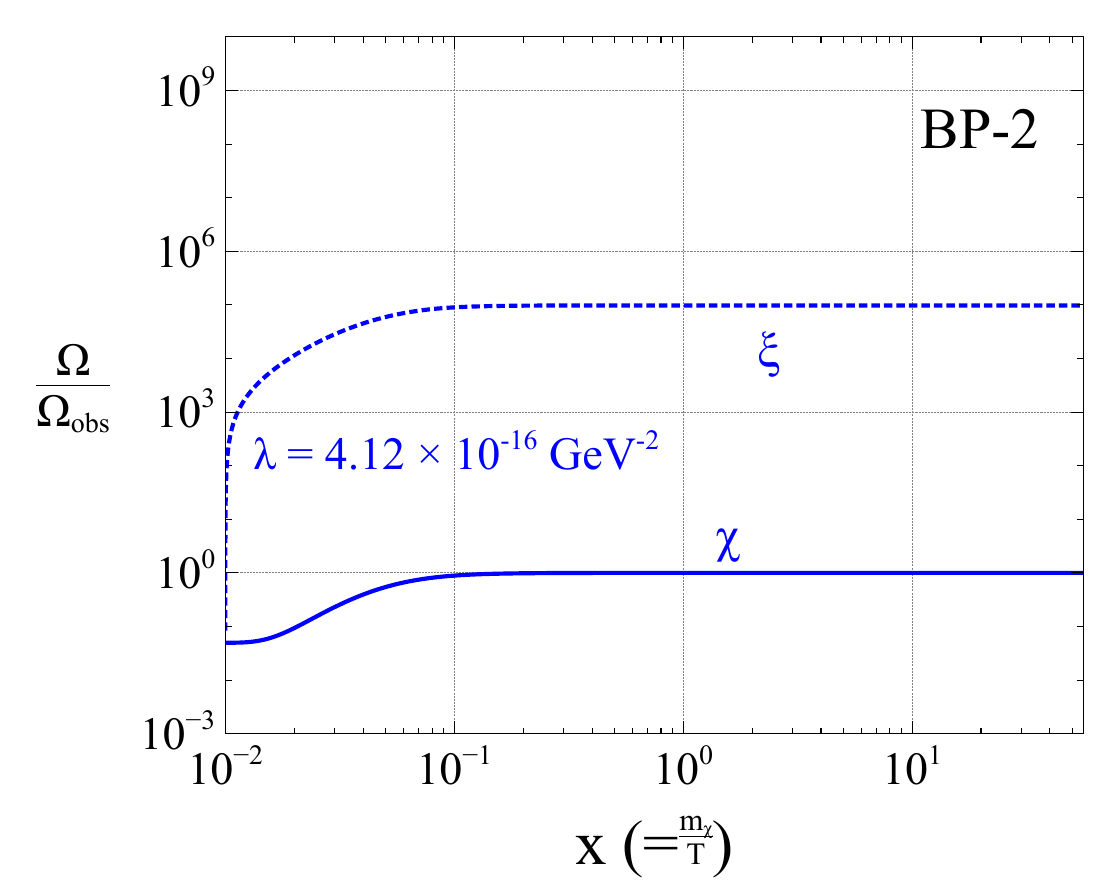}
    \caption{Evolution of $\frac{\Omega}{\Omega_{\text{obs}}}$ as functions of $x (=m_\chi/T)$ for $\chi$ (solid) and $\xi$ (dashed) for scenarios BP-1 (left) and BP-2 (right) under Case-II.
    Results for both freeze-out (red) and freeze-in (blue) scenarios are shown. The value of $\lambda_2$ that results in the observed relic abundance of the DM for the freeze-out (freeze-in) scenario is indicated in red (blue) color.}
    \label{fig:relic_case_II}
\end{figure}

In contrast, a much smaller value of the $\lambda_2$ suffices for the freeze-in mechanism to work. For such a small $\lambda_2$, the DS is unable to reach a chemical equilibrium because of an insufficient interaction rate inside the DS. Here also, the process driven by $\lambda_1$ (i.e., $\overline{\xi^c}\xi \to \overline{\xi} \chi$) will be missing and the process $\overline{\xi} \xi \to \overline{\chi} \chi$, driven by $\lambda_2$, would only contribute in achieving the right DM relic abundance via freeze-in. In Fig.~\ref{fig:relic_case_II}, we present the evolutions of the number densities ($\Omega_i/\Omega_{\text{obs}}|_{i=\chi, \xi}$) with $x = \frac{m_\chi}{T}$ 
for the scenarios BP-1 (left plot) and BP-2 (right plot). 

As can be seen from the left plot of Fig.~\ref{fig:relic_case_II}, out of the two chosen values of $\lambda_2$, the larger (smaller) one leads to, as expected, a freeze-out (freeze-in) of the DM ($\chi$) indicated by the solid red (blue) lines. It also shows that as the DM freezes out, $\xi$ attains a constant number density via freeze-in. It may, however, be noted that 
 $\xi$'s would subsequently decay to SM leptons well before the BBN.
The indicated values of $\lambda_2$, though primarily chosen to reproduce the right relic abundance via the respective mechanisms, are also checked to be consistent with various relevant theoretical and experimental constraints as elucidated later in this section by referring to Fig.~\ref{fig:scan}.

It is further noted that  
for freeze-in to work in Case-II, we require a slightly larger value of $\lambda_2$ when compared to what is needed in scenario BP-1 under Case-I. This is something expected given the contribution to the production of the DM from $\overline{\xi^c}\xi \to \overline{\xi} \chi$ is negligible (since $\lambda_1 = 0$) and hence $\overline{\xi}\xi \to \overline{\chi} \chi$ would be the only contributing process. Thus, a slightly larger value of $\lambda_2$ is required to compensate for the small deficit in the DM number density (which has been previously generated via the process $\overline{\xi^c}\xi \to \overline{\xi} \chi$ in Case-I) such that the DM could achieve the observed relic abundance.

The plot on the right in Fig.~\ref{fig:relic_case_II} simply tells us that while remaining consistent with all relevant constraints the only way in which the DM could attain the right relic abundance in BP-2 under Case-II is via freeze-in. The reasons behind this are already pointed out earlier in this subsection.

In Table~\ref{tab:final_result}, we present a comparative summary of the possibilities in the two cases we have discussed by referring to the chosen benchmark points and indicating the dominant processes responsible for DM (relic) production that contribute to the mechanism of freeze-out and/or freeze-in.
\begin{table}[ht]
	\begin{center}
		\begin{tabular}{| c | c | c | c | c | c |}
			\hline
			Case & Benchmark & Mechanism & $\lambda$ ($\lambda_2$) in Case-I(II)  & Processes & Possibility  \\
            & & & [GeV$^{-2}$] & &  \\
			\hline
			\multirow{2}{*}{Case-I} & \multirow{2}{*}{BP-1} 
            & Freeze-out & $ - $& \rule{0pt}{1.5em} $\overline{\xi} \chi \to \overline{\xi^c}\xi$ & $\times$ \\ \cline{3-6}
            & & Freeze-in & \rule{0pt}{1.5em} $1.39 \times 10^{-16}$ & $\overline{\xi^c}\xi \to \overline{\xi} \chi$, $\overline{\xi}\xi \to \overline{\chi} \chi$ & $\checkmark$ \\ 
            \hline
            \multirow{4}{*}{Case-II} & \multirow{2}{*}{BP-1} 
            & Freeze-out & \rule{0pt}{1.5em} $4.03 \times 10^{-7}$ & $\overline{\chi} \chi \to \overline{\xi} \xi$ & $\checkmark$ \\ \cline{3-6}
            & & Freeze-in & \rule{0pt}{1.5em} $1.82 \times 10^{-16} $& $\overline{\xi}\xi \to \overline{\chi} \chi$ & $\checkmark$ \\ \cline{2-6}
            & \multirow{2}{*}{BP-2} 
            & Freeze-out & \rule{0pt}{1.5em} $ - $ & $\overline{\chi} \chi \to \overline{\xi} \xi$ & $\times$ \\ \cline{3-6}
            & & Freeze-in & \rule{0pt}{1.5em} $4.12 \times 10^{-16}$ & $\overline{\xi}\xi \to \overline{\chi} \chi$ & $\checkmark$ \\ \hline
		\end{tabular}
		\caption{Possibilities of freeze-out and/or freeze-in as means of yielding the observed DM relic abundance and the processes contributing to them in Case-I and Case-II for the chosen benchmark points.}
		\label{tab:final_result}
	\end{center}
\end{table}

Before we end this discussion, we carry out an investigation on the effect of the ratio $m_\chi / m_\xi$ on such a study. Towards that, we first consider the same $C_\tau$ and $m_\xi$ as in BP-1 (for which both freeze-out and freeze-in are viable) under Case-II but consider $m_\chi = 2 \times m_\xi$ in place of $m_\chi= 1.1 \times m_\xi$ that we have used so far. We find that, for the freeze-out (freeze-in) mechanism to work, one needs to consider a smaller value of $\lambda_2$ to avoid an under-abundant (over-abundant) DM relic. Note that a smaller value of $\lambda_2$ is only constrained by the requirement of generating the observed DM relic while 
its larger values attract bounds from the demand on its perturbativity and the indirect DM-detection experiments searching for diffused $\gamma$-rays. Hence, the smaller the value of $m_\chi / m_\xi$ is, the more conservative the scenario becomes from the viewpoint of perturbativity and
bounds coming from an indirect DM-detection experiment like the Fermi-LAT. Hence our choice of $m_\chi /m_\xi=1.1$.
Furthermore, the variation of the required value of
$\lambda(=\lambda_1=\lambda_2)$ with an increasing DM mass is found to be very similar to that in Case-I.
%
\subsubsection{Indirect detection of the DM in \texorpdfstring{$\gamma$}{gamma}-ray searches}
\label{indirectSearch}
In the scenario presented in this work, prompt photons ($\gamma$) can be generated from one-step cascades of a pair of $\xi$'s produced in the mutual annihilation of the DM particles. Schematically, this is given by
$\overline{\chi} \chi \to \overline{\xi} \xi$, followed by the decays $\xi_R \to \overline{\tau_R}\nu_{\tau} \mu_L, \xi^c_R \to \overline{\tau_R}\nu_{\tau} \tau_L$ which are governed by the portal coupling $C_\tau$.  The prompt photons have their origins in the QED and electroweak bremsstrahlung off the charged leptons. The differential photon flux originating from such a cascade is given by~\cite{Bringmann:2012ez, Essig:2013goa, Slatyer:2017sev}
\begin{equation}
\label{eq:diff_flux}
\dfrac{d \Phi}{d E_\gamma \Delta \Omega}
=
\dfrac{\langle \sigma v \rangle _{\bar{\chi}\chi\to \bar{\xi} \xi}}{16\pi m_\chi^2}
\rho_\odot^2 R_\odot \bar{J}_{\rm ann}
\sum_{l} {\rm Br} (\xi \to l + \nu ) \dfrac{d N^l_\gamma}{d E_\gamma}\,\,,
\end{equation}
where $\langle \sigma v \rangle _{\bar{\chi}\chi\to \bar{\xi} \xi}$, ${\rm Br} (\xi \to l + \nu )$ and $m_\chi$ are the model-dependent quantities. Here, we have adopted the Navarro–Frenk–White (NFW) density profile for the DM~\cite{Navarro:1995iw, Navarro:1996gj} which sets the DM mass-density $\rho_\odot = 0.3 \,\text{GeV. cm}^{-3}$ at the solar location while $R_\odot$ is the distance vector between the galactic center and the Sun. The term $\bar{J}_{\rm ann}$ is completely determined by the distribution of the DM mass density $\rho(r)$ and is defined as
\begin{align}
   \bar{J}_{\rm ann} &= \frac{1}{\Delta\Omega} \int d \Omega \int \frac{d\ell}{R_\odot}  \left(\frac{\rho[r (\ell,\psi)]}{\rho_\odot}\right)^2,
\end{align}
where, `$r$' is the distance of the source under survey from the galactic center, $\ell$ is its distance from the observer along the line of sight, $\psi$ is the angle between the line of sight and $R_\odot$, $ \text{Br}(\xi \to l + \nu)$ is the relevant decay branching fractions of $\xi$ to charged leptons. The quantity  $\dfrac{d N^l_\gamma}{d E_\gamma}$, i.e., the photon energy spectrum is obtained by using the publicly available package  \texttt{PPPC4DMID}\,\cite{Cirelli:2010xx}. 
\begin{center}
	\begin{figure}[t]
\includegraphics[height=6cm,width=7.8cm]{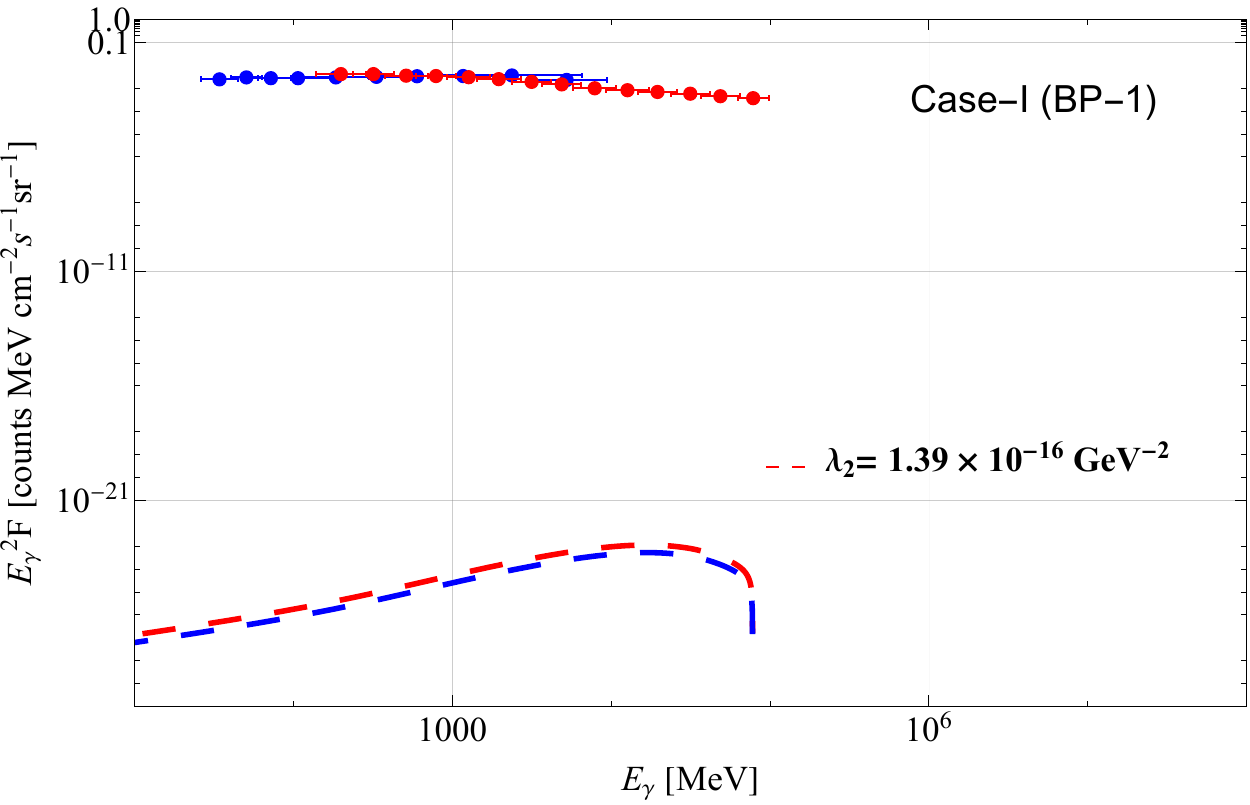}
\hspace{0.5cm}
\includegraphics[height=6cm,width=7.8cm]{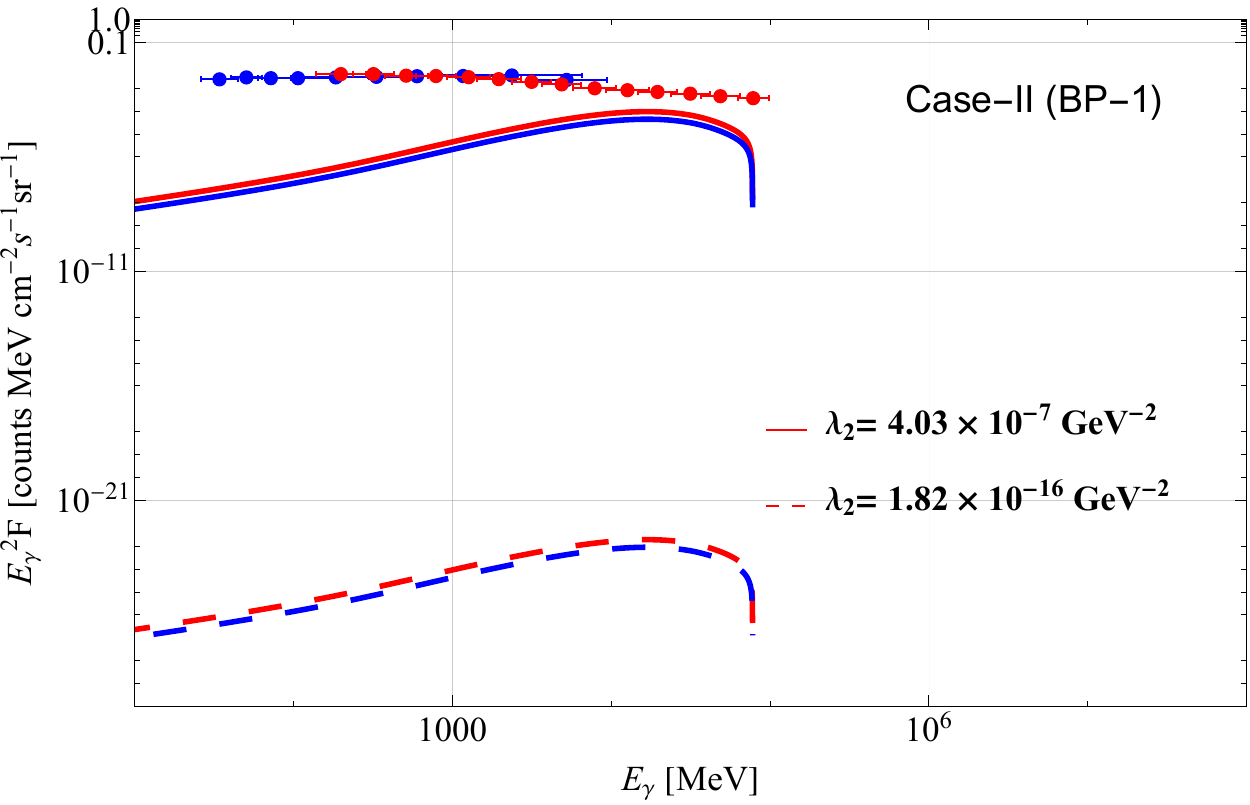}
\caption{Variations of differential photon flux, $E_\gamma^2 F$ (where $F=\frac{d\Phi}{dE_\gamma}$) as a function of $E_\gamma$
for Case-I (left) and Case-II (right), in scenarios with freeze-out (solid line) and freeze-in (dashed line). In the top part of each plot shown are the data points from the Fermi-LAT~\cite{Fermi-LAT:2012edv, Fermi-LAT:2015sau} (in red) and the EGRET~\cite{Strong:2003ey, Strong:2004de} (in blue) experiments. The corresponding values of $\lambda$ are as indicated in Fig.~\ref{fig:line_plot_1_temp} and Fig.~\ref{fig:relic_case_II}.}
\label{fig:indirect_line} 
	\end{figure}
\end{center}

We now move on to study the relevant photon spectra numerically and check how those compare with the latest experimental results (limits) from various experiments (like Fermi-LAT~\cite{Fermi-LAT:2012edv, Fermi-LAT:2015sau}, EGRET~\cite{Strong:2003ey, Strong:2004de}, in particular) that carry out indirect searches for the DM. In Fig~\ref{fig:indirect_line}, we present the energy spectra of photon flux, projected (at the bottom) for the experiments Fermi-LAT  (red) and EGRET  (blue) and compare them with the corresponding observations (at the top).  The left (right) plot refers to the scenario BP-1 under Case-I (Case-II) where the indicated values of $\lambda$ ($\lambda_2$) ensure a stable DM with the observed relic abundance (as discussed in section~\ref{sec:numerical_analysis}). 
In both cases, the smaller value of $\lambda$ ($\lambda_2$) (previously shown in blue color in Fig.~\ref{fig:line_plot_1_temp} and Fig.~\ref{fig:relic_case_II}) corresponds to the freeze-in scenario (shown in dashed lines), and the photon flux remains far below the level of current sensitivity of both Fermi-LAT and EGRET over the relevant range of $E_\gamma$.

Note that the left plot of Fig.~\ref{fig:indirect_line} tells us that the prospects of detecting the DM at the said experiments, via the $\gamma$- ray spectra, in Case-I, is abysmal.
The situation is no different in Case-II, for the freeze-in scenario, as can be gleaned from the right plot of Fig.~\ref{fig:indirect_line}. However, the projected photon fluxes at these experiments for a larger value of $\lambda_2$ (shown by solid lines) that correspond to the freeze-out scenario in BP-1 under Case-II (presented in red color in the left plot of Fig.~\ref{fig:relic_case_II}) come much closer to the current observations (sensitivities).
This is because a larger value of $\lambda_2$ enhances the key annihilation rate of the DM ($\chi$) to $\xi$, the photons contributing to the flux being subsequently radiated off the leptons to which these $\xi$'s decay to.
No doubt the situation is much better compared to the freeze-in case. However, it is to be closely checked if an increase in sensitivity by one to two orders of magnitude is possible in the future so that the flux projected here could be experimentally meaningful. 
%
\subsection{The viable region of the parameter space}
\label{sec:final_overview}
In this section, we delineate the viable region of the parameter space of the broad scenario under consideration by including the relevant theoretical and experimental constraints in our analysis. The theoretical constraints include the ones coming from the requirements of the VS and the DS to be out of mutual thermal equilibrium, and the perturbativity of the interaction strengths in the form of $\lambda_2$'s while the experimental ones refer to those required to obtain a successful BBN and not to exceed the photon flux as observed by the Fermi-LAT experiment in its bid to detect the DM indirectly. The necessary discussions on these constraints can be found under sections \ref{Thermal_equilibrium_region}, \ref{BBN} and \ref{indirectSearch}. In addition, we require compliance with the observed relic abundance of the DM. 

For the purpose at hand, we choose to work under Case-II. This is because, for Case-I, freeze-in is the only possible relic 
production mechanism which already requires a rather small 
$\lambda (=\lambda_1=\lambda_2)$ and hence far less sensitive 
to the current $\gamma$-ray searches, as can be seen from Fig.~\ref{fig:indirect_line}.
Thus, the free parameters that enter the study are $m_\chi$, $m_\xi$, $C_\tau$ and $\lambda_2$.

To understand the interplay of these parameters on the DM 
relic abundance and also on the production mechanisms of the 
DM, we thus scan over the suitable region of the parameter space under Case-II by retaining $m_\chi/m_\xi=1.1$. Such a region 
is so chosen that it complies with all the constraints 
mentioned above. In particular, the need for the VS and the 
DS to remain out of thermal equilibrium (as discussed 
previously in section~\ref{Thermal_equilibrium_region}) 
pushes one to small values of $C_\tau$. On the other hand, 
smaller $C_\tau$'s are constrained by requiring a successful BBN, as discussed in section~\ref{BBN}.

\begin{figure}[t]
 \begin{center}
		\includegraphics[height=6.8cm,width=7.8cm]{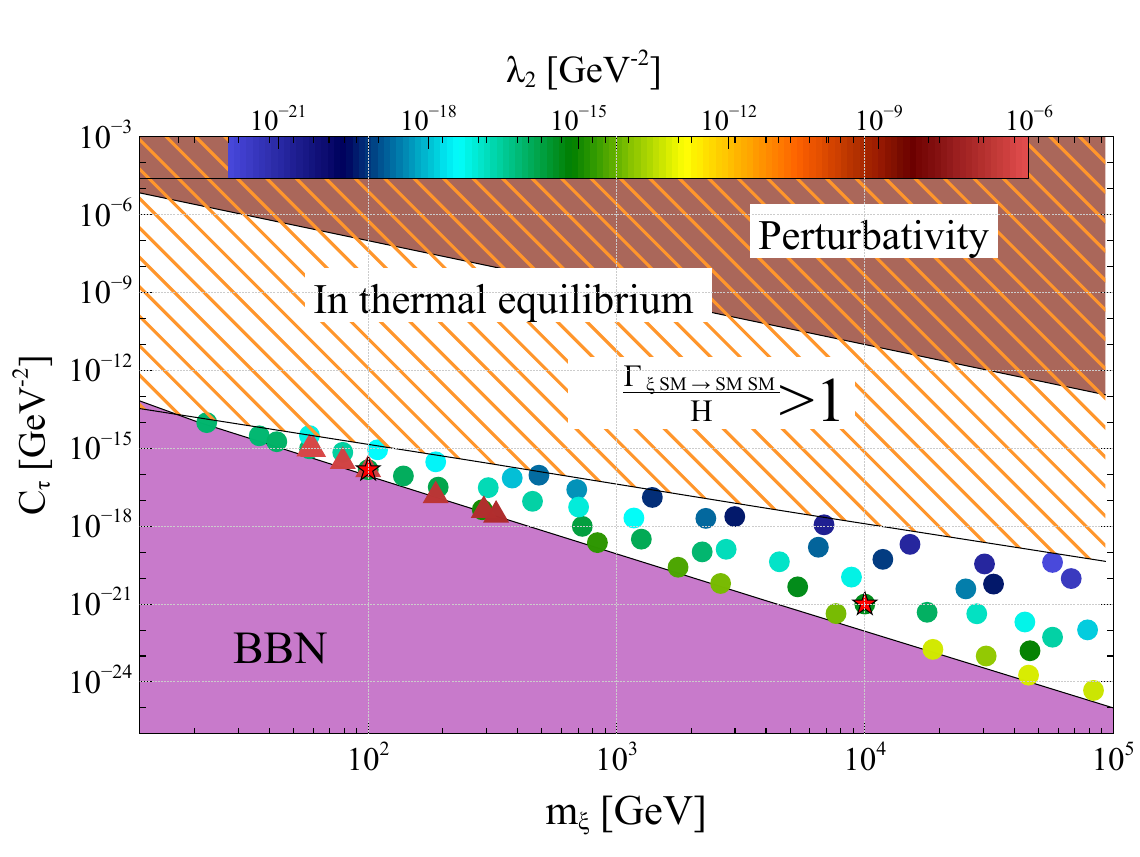}
  \hspace{0.5cm}
            \includegraphics[height=6.8cm,width=7.8cm]{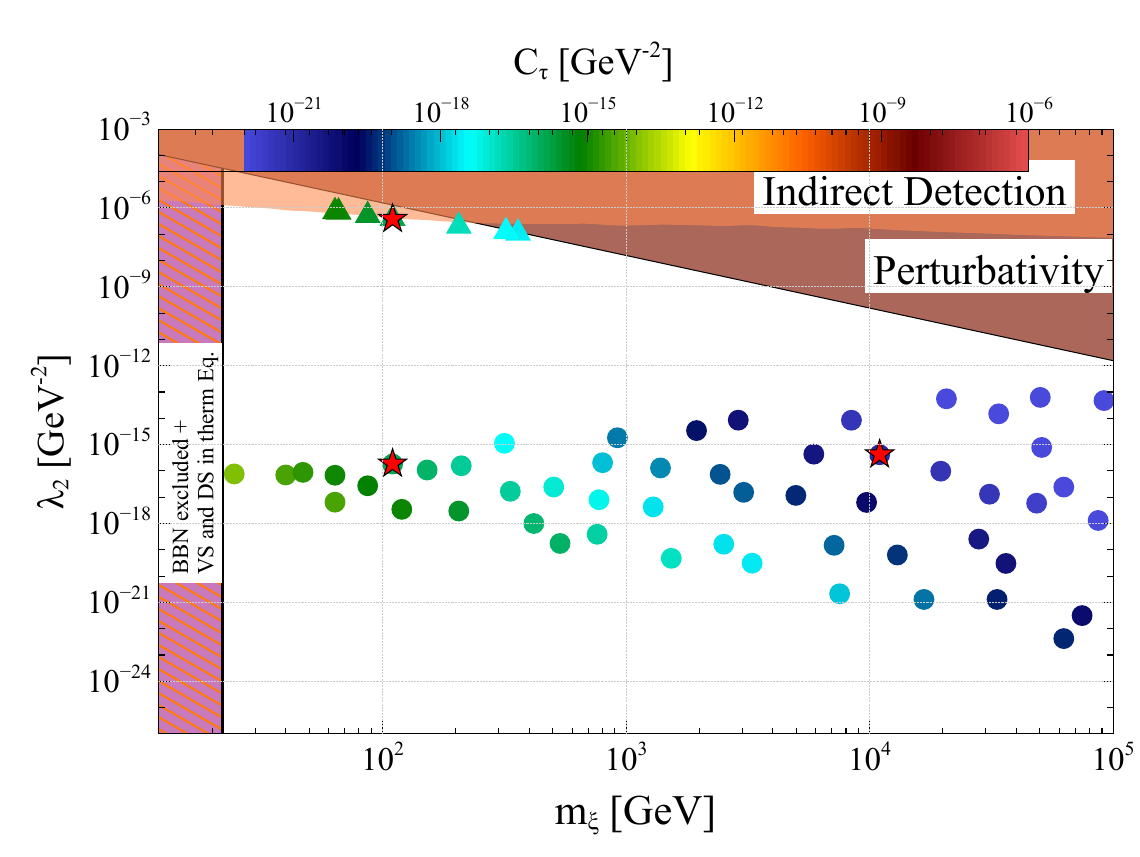}
		\caption{Regions in the $m_\xi-C_\tau$ plane (left) and $m_\chi-\lambda_2$ plane (right) constrained by requiring a successful BBN (in purple) and perturbativity (in brown) of $C_\tau$ (left) and $\lambda_2$ (right). The colored points scattered over the white regions represent parameter values which, in addition, comply with the observed DM relic abundance, the colors being indicative of the values of $\lambda_2$ (left) and $C_\tau$ (right), as shown in the respective palettes. 
  The freeze-out (filled triangles) and freeze-in (filled circles) processes via which the observed DM abundance is achieved are also presented. The asterisks (in pink) stand for the two benchmark points that we use for the numerical study. In the left plot, the region over which the VS and DS are in thermal equilibrium is delineated by orange hashed lines. In the right plot, the horizontal band in light brown is excluded by the Fermi-LAT~\cite{Fermi-LAT:2012edv,Fermi-LAT:2015sau} data. 
       }
		\label{fig:scan}
  \end{center}
\end{figure}

In Fig.~\ref{fig:scan}, we first delineate (in colors) the regions in the
relevant parameter planes that are constrained by various
theoretical and experimental consideration as discussed earlier. The left-over regions (in white) are then populated by points that satisfy the observed relic abundance (within $1\sigma$ of its observed central value). In either case, the color palette reflects the value of the remaining parameter ($\lambda_2$ or $C_\tau$) that varies simultaneously. The particular mechanism that leads to the relic abundance is indicated by a solid triangle
(freeze-out) and a solid circle (freeze-in), respectively.
The specific scenarios we considered under Case-II in the previous section are marked by solid asterisks.

For the plot on the left in Fig.~\ref{fig:scan}, we choose the plane
$m_\xi-C_\tau$ while the variation of $\lambda_2$ is presented with the help of the color palette. The choice of the variable $m_\xi$ in the horizontal axis is motivated by the fact that the appearance (or absence, which is what we try to explore and ensure in our study) of a thermal equilibrium between the VS and the DS is directly connected to $m_\xi$, rather than $m_\chi$. The phenomenon further depends on $C_\tau$. 

The same is true for the case of BBN. 
As discussed in section~\ref{Thermal_equilibrium_region}, with increasing $m_\xi$, $C_\tau$ has to drop if the VS and the DS have to remain out of thermal equilibrium. In Fig~\ref{fig:scan}, we highlight the region (purple-shaded) where the smallness of $C_\tau$ makes $\xi$ decay at a later time, i.e., at a temperature below 3~MeV, thus spoiling the prospect of a successful BBN. As can be gleaned from the left plot of Fig~\ref{fig:scan} 
(tip of the white wedge), the two aforesaid constraints, in conjunction, put a simultaneous lower bound on $m_\xi$ by excluding $m_\xi< 20 \, \text{GeV}$ and an upper bound on $C_\tau$ by excluding $C_\tau > 10^{-14} \, \text{GeV}^{-2}$
over the requirement of the VS and the DS to be thermally off-equilibrium. 

This plot further shows that the wedge-shaped white region in the $m_\xi-C_\tau$ plane that survives various constraints is moderately thin with smaller $m_\xi$ (amounting to $m_\chi \lesssim 20$ GeV) being practically ruled out. 
We further note that a freeze-out of the DM could only happen in a restricted region with not too heavy a DM ($\lesssim 400$ GeV). For larger DM masses, freeze-in is found to be the only possibility. This is because of the low values of $\lambda_2$ that we get confined to as the upper bound on it coming from its perturbativity gets drastically lowered as the DM mass increases (see section~\ref{Model}). This dependence is going to show up rather neatly in the right plot of Fig.~\ref{fig:scan}. It is evident from colors taken by the scattered points and the gradient of the color palette that, for a particular $m_\xi$, as $C_\tau$ increases the required value of $\lambda_2$ that satisfies the observed DM abundance decreases. This is because, for a particular $m_\xi$, a larger $C_\tau$ would imply a large number of $\xi$ and hence the production rate of the DM via the freeze-in mechanism will also get enhanced. Hence, to satisfy the observed relic abundance, we would require a smaller $\lambda_2$.
In the right plot of Fig.~\ref{fig:scan}, we present the same points that are presented in the left plot but in the $m_\chi-\lambda_2$ plane with $C_\tau$ now being presented via the color palette. The chosen plane suits well the simultaneous presentation of the relevant constraints including the one from the indirect detection of $\gamma$-ray by the Fermi-LAT experiment
along with that for the DM relic abundance (and the mechanisms via which this is achieved) in a single frame.
Here we present how the $\gamma$-ray flux observed by the Fermi-LAT experiment~\cite{Fermi-LAT:2012edv, Fermi-LAT:2015sau} (the one from the EGRET experiment being less severe in the region we are interested in) restricts the $m_\chi-\lambda_2$ plane. 
To find this out, we perform a small scan over the parameters \{$\lambda_2$, $m_\chi$\} while setting $m_\xi = \frac{m_\chi}{1.1}$. For some given sets of values for $\{m_\chi,\lambda\}$\footnote{The differential photon flux is proportional to $\text{Br}(\xi\to l+\nu)$ as shown in Eq.~(\ref{eq:diff_flux}). The quantity is insensitive to $C_\tau$ as it gets canceled out in the definition of branching fraction.}, 
  we then estimate the differential photon flux of Eq.~(\ref{eq:diff_flux}) in $E_\gamma$ and ensure that the value of the flux thus obtained does not exceed its upper bound as set by Fermi-LAT~\cite{Fermi-LAT:2012edv, Fermi-LAT:2015sau}. This then rules out a region in the $m_\chi - \lambda_2$ plane which is delineated in the figure in a light-brown shade.

The region that is ruled out by demanding perturbativity of $\lambda_2$ is delineated by a deep brown color. Also shown is a vertical band (in orange hash over a purple region) at small values of $m_\chi$ which arises from the simultaneous imposition of the requirement of a successful BBN and the need for the VS and the DS to be off-equilibrium as discussed in the context of the left plot. In fact, this is a map in the current plane of a robust observation made from the plot on the left that $m_\chi \lesssim 20$ GeV is ruled out in such a numerical setup.

As may be seen, the points are now clearly grouped in two regions: for the few 
points having larger values of $\lambda_2$ and $m_\chi \lesssim 400$ GeV, the 
observed DM relic abundance is achieved, as expected, via the freeze-out (solid 
triangles) route. On the other hand, for lower values of $\lambda_2$, the 
freeze-in (solid circles) mechanism yields the correct DM relic abundance over 
the entire range of $m_\chi$ presented in the plot.

It is to be noted that the freeze-out route seems to be somewhat fine-tuned as 
the corresponding points appear at the edge of viability when seen from the 
angle of perturbativity of $\lambda_2$ and the constraints imposed on it from 
an indirect DM detection experiment like the Fermi-LAT. The fine-tuned situation 
arises since a somewhat lower value of $\lambda_2$ (with other parameters kept 
fixed) would lead to an over-abundant DM via freeze-out. The only possibility 
to get back the right DM relic is then by opting for a smaller $C_\tau$ which 
would suppress the production of $\xi$ at the early Universe leading to a 
smaller effective equilibrium number density ($n_{\chi}^{\text{eq,eff}}$) of 
the DM (see Eq.~\ref{Eq:effEqDM}). Consequently, a reduced annihilation rate 
of the DM resulting from a lower value of $\lambda_2$ could have been 
sufficient to optimally annihilate the lesser number of the DM given by the 
suppressed value of $n_{\chi}^{\text{eq,eff}}$ thus retrieving the observed 
value of DM relic abundance.  However, it is clear from the left plot of 
Fig.~\ref{fig:scan} that such a smaller value of $C_\tau$ is ruled out by 
requiring a successful BBN. Hence the absence of any viable point at a lower 
value of $\lambda_2$ which results in such a fine-tuned situation.  

As for the larger set of points with lower values of $\lambda_2$ for which observed DM relic abundance is achieved via freeze-in, the following patterns cannot be missed.
\begin{itemize}
\item
For any given value of $m_\chi$, as $\lambda_2$ increases, $C_\tau$ is found to decrease. This is because with increasing $\lambda_2$, the annihilation rate of $\xi$ increases (via $\bar{\xi} \xi \to \bar{\chi} \chi$) and consequently, we end up with an over-abundant DM. Hence, to obtain the DM relic abundance within the experimentally observed range, a lowering of $C_\tau$ so that the number density of $\xi$ initially produced from the VS (via the generic process SM SM $\to$ SM $\xi$, see Eqs.~\ref{eq:charged_higgs_process} and \ref{eq:neutral_higgs_process}) is reduced. However, lowering of $C_\tau$ is soon confronted by the requirement of a successful BBN. This is reflected by the absence of any viable point beyond a certain value of $\lambda_2$ depending on the value of $m_\chi$.
Conversely, $\lambda_2$ can not indefinitely decrease as that would make the DM underabundant via freeze-in.
To compensate for this, one could instead increase $C_\tau$. However, this would soon restore the thermal equilibrium between the VS and the DS (see section~\ref{Thermal_equilibrium_region}) thus spoiling the very basic
requirement for such a secluded scenario to work. 
\item
For any given $\lambda_2$, as $m_\chi$ ($m_\xi$) increases, $C_\tau$ is found to decrease, as can be gleaned from the color palette. This is because, with increasing $m_\chi$, the reaction rate of the process $\bar{\xi} \xi \to \bar{\chi} \chi$ that controls the freeze-in mechanism gets enhanced (as it is driven by a four-fermion operator which makes $\langle \sigma v \rangle$ vary like $m_\chi^2$) thus resulting in an over-abundant DM. A smaller $C_\tau$
could bring down the abundance to its observed value by lowering $n_\xi$.
The plot, however, displays a clear trend that a further increase in $m_\chi$ would soon attract the perturbative bound on $\lambda_2$ for its larger values.
\end{itemize}
The discussions above reveal that, in the case under consideration, the freeze-out mechanism that results in the right amount of DM relic requires a relatively large $\lambda_2 ({\cal O}(10^{-7} \, \textrm{GeV$^{-2}$}))$. Its viability, however, is somewhat strained due to looming constraints from the theory (perturbativity of $\lambda_2$) and the Fermi-LAT experiment. The discussions also shed light on how intimately the observed DM abundance achieved via the freeze-in mechanism is connected to the production of $\xi$ from the SM bath at the early Universe and its annihilation rate. 

\subsection{Constraints from Lepton-Flavour Universality Violation}
\label{subsec:LFUV}
\begin{figure}[t]
\begin{center}
\includegraphics[width=0.3\linewidth]{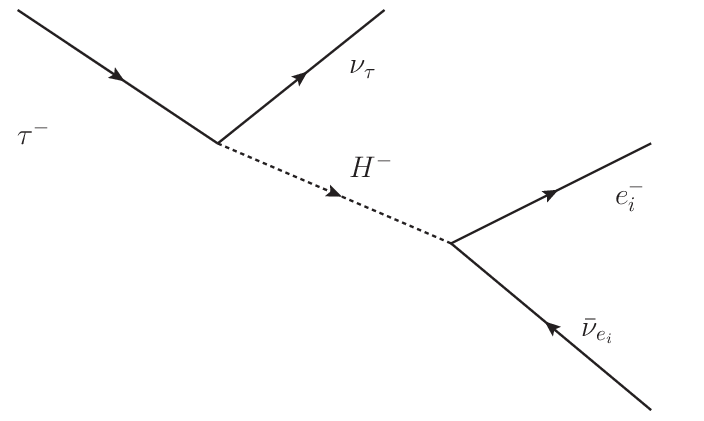}
    \caption{Feynman diagram responsible for LFUV contribution from BSM, where $i = \mu, \tau$.}
    \label{fig:placeholder}
    \end{center}
\end{figure}

Apart from the astrophysical and cosmological constraints on the model parameter space, we have also examined the limits arising from LFUV.
In the SM, the decays $\tau \to \mu \bar{\nu}_\mu \nu_\tau$ and $\tau \to e \bar{\nu}_e \nu_\tau$ proceed through a virtual $W^\pm$ boson.
In our framework, integrating out the heavy charged Higgs $H^\pm$ induces additional dimension-6 operators that contribute to these processes (as shown in fig.~\ref{fig:placeholder}). Since the resulting decay amplitudes depend on the Yukawa interactions of the $\mu$, $\tau$, and $e$, they modify the LFUV observable $(g_\mu/g_e)$, which is constrained by $\tau$-decay rate and its leptonic branching fractions. In lepton specific 2HDM, at tree level, $(g_\mu/g_e)$ can be expressed as~\cite{Logan:2009uf}
\begin{align}
    \frac{g_\mu^2}{g_e^2} = \frac{
1 + m_\mu^2 m_\tau^2 \tan^4\!\beta \,/\, (4 M_{H^\pm}^4)
\;-\;
\left( 2 m_\mu^2 \tan^2\!\beta \,/\, M_{H^\pm}^2 \right)
\, g\!\left( \frac{m_\mu^2}{m_\tau^2} \right)
\big/ f\!\left( \frac{m_\mu^2}{m_\tau^2} \right)
}{
1 + m_e^2 m_\tau^2 \tan^4\!\beta \,/\, (4 M_{H^\pm}^4)
\;-\;
\left( 2 m_e^2 \tan^2\!\beta \,/\, M_{H^\pm}^2 \right)
\, g\!\left( \frac{m_e^2}{m_\tau^2} \right)
\big/ f\!\left( \frac{m_e^2}{m_\tau^2} \right)
}\,,
\end{align}
where, $f(x) = 1 - 8x + 8x^{3} - x^{4} - 12x^{2}\ln x,\quad
g(x) = 1 + 9x - 9x^{2} - x^{3} + 6x(1+x)\ln x$.
The constraint on $(g_\mu/g_e) = 1.0018 \pm 0.0014$ is given by HFLAV group~\cite{HFLAV:2014fzu}.
As discussed in Ref.~\cite{Logan:2009uf}, these considerations make the following windows of $m_{H^{\pm}}$ viable: 
\begin{align}
 (0.60\, \tan{\beta})\, \text{GeV} \leq m_{H^{\pm}}\leq (0.66\, \tan{\beta})\, \text{GeV}\,,  \quad m_{H^\pm} \geq (3.15\,\tan\beta) \,\text{GeV} \,.
    \label{eq:mh_lfuv}
\end{align}
As discussed already, in our scenario, the relevant region of parameter space should have large $m_{H^{\pm}} (>100\, m_\chi)$ . 
Since we are working in an EFT scenario, this puts a lower limit on the effective scale $\Lambda$. Therefore by using the definition of $C_{\tau}$ (in Eq.~\ref{eq:ctau}),
Eq.~\ref{eq:mh_lfuv} can be expressed in terms of our model parameters as
\begin{align*}
    C_{\tau} \leq \frac{g_2 \, m_\tau\, y_4\, \cos{\beta}}{\sqrt{2}\,m_W (3.15)^2} \,,
\end{align*}
where, $g_{2}$ is the $SU(2)_L$ gauge coupling of the SM. For the maximum value of $\tan\beta$ (as given in Eq.~\ref{eq:tanbeta}), 
\begin{align}
    \frac{C_\tau}{y_4} \leq 8.2 \times 10^{-7}\,\text{GeV}^{-2} \,.
\end{align}
From Fig.~\ref{fig:scan}, the maximum contribution to the ratio $\frac{C_\tau}{y_4}$ comes from  the point $\lbrace{m_{\xi}\,,C_{\tau}\rbrace} = \lbrace{ 15.4\, \text{GeV}\,, 10^{-14}\, \text{GeV}^{-2} \rbrace}$. Using the definition of $C_{\tau}$, we recalculate the ratio $\frac{C_\tau}{y_4} \simeq 3.9 \times 10^{-9}\,\mathrm{GeV}^{-2}$ (for maximum $\tan{\beta})$.
Thus we can see that the ratio 
lies well below the conservative LFUV bound, ensuring that the model safely satisfies the experimental constraint on $\frac{g_\mu}{g_e}$ as mentioned above.
%
%
\section{Summary and Conclusions}
\label{summary}
In this work, we present a theoretical description and the
phenomenology of a secluded dark matter having its origin in a
setup motivated by an effective field theory. We have explored
the type-X 2HDM with a global $U(1)_{\mu - \tau}$ symmetry extended
by two fermionic fields $\chi$ (the DM) and $\xi$ (the mediator) 
and a scalar field `$S$', all belonging to a secluded dark
sector. The choice of such a scenario is motivated by the
possibilities that it allows for an initial thermal production
of the dark sector particle, $\xi$, from the visible (SM) sector
and the mutual conversion processes involving $\xi$ and $\chi$ 
within the dark sector working in tandem to yield the observed
DM relic abundance. The former set of processes is predominantly 
controlled by various dim-6 four-Fermi (contact interaction) 
operators that we have constructed by integrating out the 
heavier Higgs bosons ($H^{\pm}$, $H$ and $A$), with their 
strengths being given by the portal coupling parameter $C_\tau$. The 
latter ones are constructed by integrating out the heavy
dark-sector scalar excitation ($S$) whose strengths are 
parametrized by the DS couplings $\lambda_{1,2}$. Further, on 
the cosmology front, the scenario is able to ensure a successful 
BBN front and, on the astrophysics front, could also be 
sensitive to $\gamma$-rays that are searched for as indirect 
probes to the DM in an experiment like the Fermi-LAT.

We have identified the region in the $m_\xi - C_\tau$ plane 
where the VS and the DS are not in thermal equilibrium which is 
characterized by a relatively small $C_\tau$. However, the 
requirement of a successful BBN in the early Universe places a 
lower bound on $C_\tau$, for any given value of $m_\xi$ (and 
hence $m_\chi$, the mass of the DM particle). On the other hand, 
the stability of the DM ($\chi$) requires a small enough DS 
coupling $\lambda_1$. Furthermore, we have ensured the 
perturbativity of both $C_\tau$ and $\lambda_{2}$.

With all these considerations in place, we have carried out a 
comprehensive numerical analysis of the scenario by solving the 
appropriate set of coupled Boltzmann equations after incorporating all relevant production and 
annihilation processes of $\chi$ and $\xi$. Depending on the coupling 
strengths of the DS ($\lambda_1$, $\lambda_2$), we have considered two cases; 
Case-I ( $\lambda_1 = \lambda_2 = \lambda$) and Case-II ($\lambda_1 (= 0) \, \ll \lambda_2$). The salient observations of these two cases that can be made from the present analysis are as follows.
\begin{itemize}
\item
Requiring the VS and the DS to be out of mutual thermal equilibrium and simultaneously securing a successful BBN, forbids the DM mass, $m_\chi$, below around 20 GeV. This results in an absolute upper bound on $C_\tau$ ($\lesssim 10^{-14}$ GeV$^{-2}$) in the scenario under consideration. The physics behind both these requirements depends only on the portal coupling $(C_\tau)$ and the mass of the mediator ($m_\xi$), and 
is independent of the choice of $\lambda_{1,2}$, 
These requirements and the dependencies on the portal coupling and mediator mass can be considered as some general features in any scenario that can prompt non-thermal production of DM. 
\item In Case-I, the possibility of a large value of $\lambda$
($=\lambda_1=\lambda_2$) which helps attain the right DM relic via freeze-out is ruled out by the stability of the DM, $\chi$, decaying at a later time to $\xi$, via the loop-mediated process
$\chi \to \xi l \bar{l}$, even as its tree-level decay is forbidden due to the choice $m_\xi < m_\chi < 3m_\xi$. However, the correct relic abundance of the DM can still be achieved by exploiting the freeze-in mechanism, with a smaller $\lambda$, thus 
without jeopardizing the stability of the DM. This interplay of the DS couplings $\lambda_1$ and $\lambda_2$ in the production mechanism of DM is independent of the choice of the origin of DS production.
\item The setup proposed in Case-II could explain the observed DM relic abundance achieved either via freeze-out or through freeze-in within the DS depending on the strength of the interaction, $\lambda_2$ between $\chi$ and $\xi$. Here, the stability of the DM ($\chi$) is ensured by setting $\lambda_1 = 0$ which prohibits the possible loop level decay of $\chi$, $\chi \to \xi l \bar{l}$.
\item In Case-II one requires $\lambda_2= \mathcal{O}(10^{-7} \, \text{GeV}^{-2})$ for the freeze-out scenario to work. However, the adverse interplay of the model parameters, while complying simultaneously with different constraints (including indirect detection), leads to a rather fine-tuned region having smaller $m_{\xi,\chi}$ and larger $\lambda_2$. For a large DM maas, $m_\chi \gtrsim 400 \, \text{GeV}$, the scenario is constrained by the perturbativity of $\lambda_2$.
\item On the other hand, in Case-II, the freeze-in mechanism works with rather small values of $\lambda_2$, as expected, but now over a large range of $m_\chi$. The strength of the coupling parameter $\lambda_2$ (required to produce the observed DM relic abundance) has been found to be inversely proportional to the portal coupling $C_\tau$. 
\item  Note that the requirement of a larger DS coupling ($\lambda_2$) for the freeze-out scenario to work 
under Case-II would result in an enhanced $\gamma$-ray flux arising from DM annihilation. Such an enhanced $\gamma$-ray flux could easily reach the current limit of sensitivity of the Fermi-LAT experiment; hence, the scenario attracts a bound from the same flux. 

Therefore, Case-II could be valid over a larger 
volume of parameter space that conforms with the 
constraints coming from the requirement of having off-equilibrium VS and DS, as well as from
the BBN, the observed relic abundance of the DM, and those from the observed $\gamma$-ray flux in an experiment like the Fermi-LAT. 

\item The constraint on LFUV from the HFLAV experiment translates into a lower bound on the effective scale of the theory and hence results in an upper bound on the ratio $\frac{C_{\tau}}{y_4}$. The latter turns out to be far above the conservative limit of our, thus ensuring its viability.
\end{itemize}

To conclude we would like to emphasise that we focus on a minimalistic setup in which the secluded dark matter can have an EFT-based origin in a type-X 2HDM scenario. Nonetheless, we have clarified that our approach would work in a more general setup as well. In that case, qualitatively, the phenomenological aspects remain the same, albeit there could be some major quantitative implications depending on the origin of the DS. As for a future $\gamma$-ray search experiment, the Cherenkov Telescope Array (CTA)~\cite{CTAConsortium:2017dvg} would have the potential to constrain the parameter space further. This can, in turn, rule out some larger values of $\lambda_2$ thus restricting the possibility of attaining the observed DM relic in such a scenario via the freeze-out mechanism. 
 As far as the BBN is concerned, some improvements in our analysis are still possible by performing a more precise computation of the abundances of the primordial elements in the presence of late decays of an otherwise long-lived $\xi$. We leave such studies for future works.
%
\section{Acknowledgments}
AKS is supported by a postdoctoral fellowship at IOP, Bhubaneswar, India. AT would like to acknowledge the financial support provided by the Indian Association for the Cultivation of Science (IACS), Kolkata.  
%
\bibliographystyle{JHEP}
\bibliography{References}
%
\end{document}